%% file: main.tex
\def\year{2021}\relax
\documentclass[letterpaper]{article} 
\usepackage{aaai21}  
\usepackage{times}  
\usepackage{helvet} 
\usepackage{courier}  
\usepackage[hyphens]{url}  
\usepackage{graphicx} 
\usepackage{natbib}  
\usepackage{caption} 
\usepackage{amssymb}
\usepackage{enumitem}
\usepackage{listings}
\usepackage{xstring}
\usepackage{graphicx}
\usepackage{pbox}
\usepackage{subcaption}
\usepackage{epstopdf}
\usepackage{xstring}

\usepackage{balance}
\usepackage{pbox}
\usepackage{multirow}
\usepackage{enumitem}
\usepackage[normalem]{ulem}

\usepackage{multicol}

\usepackage{tabularx}
\usepackage{tcolorbox}
\usepackage{color, colortbl}
\usepackage[show]{inotes}
\usepackage{soul}
\usepackage{booktabs}
\usepackage{multirow}
\usepackage[switch]{lineno}

\usepackage{comment}

\newcommand{\deltext}[1]{} 

\title{Fighting the COVID-19 Infodemic in Social Media:\\
	A Holistic Perspective and a Call to Arms}
\author{Firoj Alam, Fahim Dalvi, Shaden Shaar,  Nadir Durrani, Hamdy Mubarak, Alex Nikolov$^*$,\\
Giovanni Da San Martino$^{**}$, Ahmed Abdelali, 
Hassan Sajjad, Kareem Darwish, Preslav Nakov\\}
\affiliations{
    Qatar Computing Research Institute, HBKU, Qatar\\
    $^*$Sofia University ``St Kliment Ohridski'', Bulgaria\\
    $^{**}$University of Padova, Italy\\
    \{fialam,faimaduddin,pnakov\}@hbku.edu.qa

}

\begin{document}
\maketitle

\begin{abstract}
With the outbreak of the COVID-19 pandemic, people turned to social media to read and to share timely information including statistics, warnings, advice, and inspirational stories. Unfortunately, alongside all this useful information, there was also a new blending of medical and political misinformation and disinformation, which gave rise to the first global infodemic. While fighting this infodemic is typically thought of in terms of factuality, the problem is much broader as malicious content includes not only fake news, rumors, and conspiracy theories, but also promotion of fake cures, panic, racism, xenophobia, and mistrust in the authorities, among others. This is a complex problem that needs a holistic approach combining the perspectives of journalists, fact-checkers, policymakers, government entities, social media platforms, and society as a whole. 
With this in mind, we define an annotation schema and detailed annotation instructions that reflect these perspectives. We further deploy a multilingual annotation platform, and we issue a \emph{call to arms} to the research community and beyond to join the fight by supporting our crowdsourcing annotation efforts. We perform initial annotations using the annotation schema, and our initial experiments demonstrated sizable improvements over the baselines.
\end{abstract}

\input{sections/introduction.tex}

\input{sections/related_work.tex}
\input{sections/annotation_platform.tex}

\input{sections/dataset.tex}
\input{sections/experiments_results.tex}
\input{sections/conclusion.tex}

\section*{Acknowledgments}

This research is part of the Tanbih project,\footnote{\url{http://tanbih.qcri.org}} developed at the Qatar Computing Research Institute, HBKU, which aims to limit the impact of ``fake news'', propaganda, and media bias by making users aware of what they are reading, thus promoting media literacy and critical thinking.

\section*{Ethics and Broader Impact}
    
Our dataset was collected from Twitter, following its terms of service. It can enable analysis of social media content, which could be of interest to practitioners, professional fact-checker, journalists, social media platforms, and policy makers. Our models can help fight the infodemic, and they could support analysis and decision making for the public good. However, they could also be misused by malicious actors. 
    
{
\small
\bibliography{bib/main,bib/checkthat19}
}

\end{document}


\linenumbers
\maketitle

\appendix
\input{sections/supplemental_material}

\newpage
\bibliographystyle{aaai21}
\bibliography{bib/main,bib/checkthat19}

%% file: sections/introduction.tex
\section{Introduction}
\label{sec:introduction}

The year 2020 brought along two remarkable events: the COVID-19 pandemic, and the resulting first global infodemic. The latter thrives in social media, which saw growing use as, due to lockdowns, working from home, and social distancing measures, people spend long hours in social media, where they find and post valuable information, big part of which is about COVID-19. Unfortunately, amidst this rapid influx of information, there was also a spread of disinformation and harmful content in general, fighting which became a matter of utmost importance. 
In particular, as the COVID-19 outbreak developed into a pandemic, the disinformation about it followed a similar exponential growth trajectory. The extent and the importance of the problem soon lead to international organizations such as the World Health Organization and the United Nations referring to it as the first global \emph{infodemic}.
Soon, a number of initiatives were launched to fight this infodemic.

The focus of these initiatives was on social media, e.g.,~building and analyzing large collections of tweet, their content, source, propagators, and spread~\citep{leng2020analysis,Medford2020.04.03.20052936,mourad2020critical,karami2021identifying,broniatowski2018weaponized}. 
Most such efforts were in line with previous work on disinformation detection, which focused almost exclusively on the factuality aspect of the problem, while ignoring the equally important potential to do harm. The COVID-19 infodemic is even more complex, as it goes beyond spreading fake news, rumors, and conspiracy theories, and extends to promote fake cures, panic, racism, xenophobia, and mistrust in the authorities, among others. This is a complex problem that needs a holistic approach combining the perspectives of journalists, fact-checkers, policymakers, government entities, social media platforms, and society. 

\begin{figure}[!htb]
\centering
\includegraphics[width=0.40\textwidth]{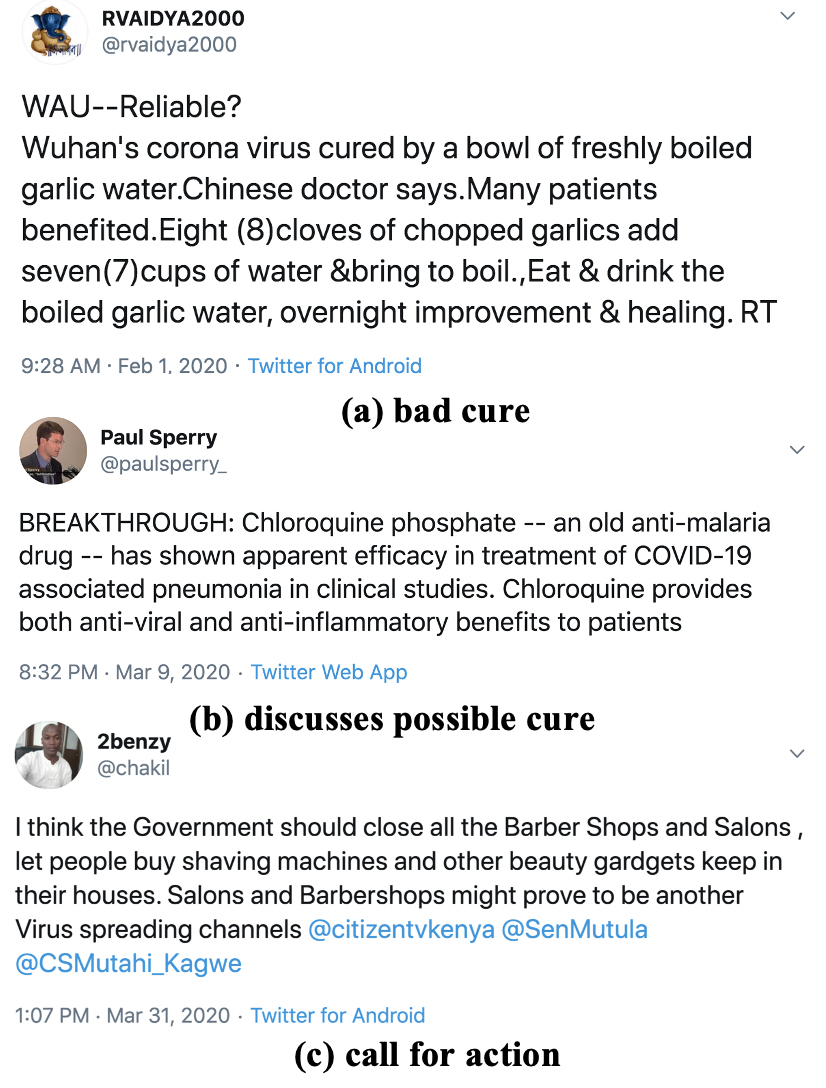}
\caption{Examples of tweets showing some issues that are important to journalists, fact-checkers, social media platforms, policy makers, government entities, and the society.}
\label{fig:tweet_example}
\end{figure}

Here we define a comprehensive annotation schema that goes beyond factuality and potential to do harm, extending to information that could be potentially useful, e.g.,~for government entities to notice or for social media to promote. Information about a possible cure for COVID-19 should get the attention of a fact-checker, and if proven false, as in the example in Figure~\ref{fig:tweet_example}(a), it should be flagged with a warning or even removed from the social media platform to prevent its further spread; it might also need a response by a public health official. However, if proven truthful, it might instead be promoted in view of the high public interest in the matter. Our annotation schema further covers some categories of good posts, including giving advice, asking a question, discussing action taken, possible cure as in Figure~\ref{fig:tweet_example}(b), or calling for action as in Figure~\ref{fig:tweet_example}(c). Such posts could be useful for journalists, policymakers, and society as a whole.

We organize the annotations around seven questions, asking whether a tweet (1)~contains a verifiable factual claim, (2)~is likely to contain false information, (3)~is of interest to the general public, (4)~is potentially harmful to a person, a company, a product, or society, (5)~requires verification by a fact-checker, (6)~poses harm to society, or (7)~requires the attention of a government entity. 
Annotating so many aspects is challenging and time-consuming. Moreover, the answer to some of the questions is subjective, which means we really need multiple annotators per example, as we have found in our preliminary manual annotations. 

Our contributions can be summarized as follows:

\begin{itemize}
    \item We develop comprehensive guidelines that combine the perspectives and the interests of journalists, fact-checkers, social media platforms, policymakers, and the society as a whole.
    \item We develop a 
    volunteer-based crowd annotation platform based on Micromappers\footnote{\label{micromapper}\url{http://micromappers.qcri.org}}, and we invite volunteers to join our annotation efforts. 
    \item We annotate initial datasets covering English and Arabic.    
    \item We present experimental results showing sizable improvements over the baselines, when using both coarse and fine-grained labels.
\end{itemize}

\section{Call to Arms}
\label{sec:call}
We invite volunteers to join our crowdsourcing annotation efforts and to label some new tweets, thus supporting the fight against the COVID-19 infodemic. We make all resulting annotations publicly available.\footnote{Our data: \url{http://doi.org/10.7910/DVN/XYK2UE}}
As of present, we focus on English
\footnote{Annotation link for English: \url{http://micromappers.qcri.org/project/covid19-tweet-labelling/}} 
and Arabic,\footnote{Annotation link for Arabic, \url{http://micromappers.qcri.org/project/covid19-arabic-tweet-labelling/}} 
but we plan to add more languages in the future.

%% file: sections/related_work.tex
\section{Related Work}

\paragraph{``Fake News'', Disinformation, and Misinformation:}
There has been a lot of interest in recent years in identifying disinformation, misinformation, and ``fake news'', which thrive in social media. The studies of \citep{Lazer1094} and \citep{Vosoughi1146} in \emph{Science} offered a general overview and discussion on the science of ``fake news'' and of the process of proliferation of true and false news online.
There have also been several interesting surveys, e.g.,~\cite{Shu:2017:FND:3137597.3137600} studied how information is disseminated and consumed in social media. Another survey by \cite{thorne-vlachos:2018:C18-1} took a fact-checking perspective on ``fake news'' and related problems. Yet another survey \cite{Li:2016:STD:2897350.2897352} covered truth discovery in general. 
Some very recent surveys focused on stance for misinformation and disinformation detection \cite{Survey:2021:Stance:Disinformation}, on automatic fact-checking to assist human fact-checkers \cite{Survey:2021:AI:Fact-Checkers}, on predicting the factuality and the bias of entire news outlets \cite{Survey:2021:Media:Factuality:Bias}, on multimodal disinformation detection \cite{Survey:2021:Multimodal:Disinformation}, and on abusive language in social media \cite{Survey:2021:Abusive:Language}.

\paragraph{Fact-Checking:} Research in this direction includes \textit{fact-checking}, i.e.,~verifying the veracity of the claim in textual and imagery content, and \textit{check-worthiness}, deciding whether a claim is worthy of investigation by a professional fact-checker. 
There have been a number of professional organizations working on \textit{fact-checking},\footnote{\url{https://en.wikipedia.org/wiki/List_of_fact-checking_websites}} and a number of dataset have been developed by the NLP research community to develop models for automatic fact-checking. Some of the larger datasets include the \emph{Liar, Liar} dataset of 12.8K claims from PolitiFact \cite{wang:2017:Short}, \emph{ClaimsKG} dataset and system~\cite{ClaimsKG} of 28K claims from 8 fact-checking organizations, the \emph{MultiFC} dataset of 38K claims from 26 fact-checking organizations \cite{augenstein-etal-2019-multifc}, and the 10K claims \emph{Truth of Various Shades} \cite{rashkin-EtAl:2017:EMNLP2017} dataset, among other smaller-size ones. 
A number of datasets have also been developed as part of shared tasks. In most cases, they did not rely on fact-checking websites, but performed their own annotation, either (a)~manually, e.g., the SemEval-2017 task 8~\cite{derczynski-EtAl:2017:SemEval} and the SemEval-2019 task 7~\cite{gorrell-etal-2019-semeval} on Determining Rumour Veracity and Support for Rumours (RumourEval), the SemEval-2019 task 8 on Fact-Checking in Community Question Answering Forums~\cite{mihaylova-etal-2019-semeval}, the CLEF 2019--2021 CheckThat! Lab \cite{CheckThat:ECIR2019,CheckThat:ECIR2020,CheckThat:ECIR2021}, which featured both English and Arabic, or (b)~using crowdsourcing, e.g.,~the FEVER 2018--2019 tasks on Fact Extraction and VERification, which focused on fact-checking made-up claims about content present in Wikipedia~\cite{thorne-etal-2019-fever2}. Datasets have also been developed for non-English languages, e.g.,~the study in \cite{baly-EtAl:2018:N18-2} developed a dataset of 402 Arabic claims extracted from Verify-SY.

\paragraph{Check-Worthiness:} As the detected claims can be large in volume, it is difficult for professional fact-checkers to check them and many claims might not be urgent to fact-check. Hence, it is important to identify check-worthy claims. 
A manually labeled dataset for check-worthiness was used in the ClaimBuster system \cite{Hassan:15}. \cite{gencheva-EtAl:2017:RANLP} developed a dataset of political debates with labels collected from fact-checking websites. This dataset was used in the ClaimRank system \cite{NAACL2018:claimrank}, and it was extended and used in the CLEF CheckThat! labs 2018-2021~\cite{clef2018checkthat:overall,CheckThat:ECIR2019,CheckThat:ECIR2020,CheckThat:ECIR2021}. 

\paragraph{Fighting the COVID-19 Infodemic}
There have been a number of COVID-19 Twitter datasets: many without labels, other using distant supervision, and very few manually annotated. Some large datasets include a multi-lingual dataset of 123M tweets~\cite{info:doi/10.2196/19273}, another one of 152M tweets \cite{Banda:2020}, a billion-scale dataset of 65 languages and 32M geo-tagged tweets~\cite{abdul2020mega}, and the GeoCoV19 dataset, consisting of 524M multilingual tweets, including 491M with GPS coordinates~\cite{Umair2020geocovid19}. 
There have also two Arabic datasets, some without manual annotations \cite{alqurashi2020large}, and some with \cite{haouari2020arcov19:rumors,ARCorona:2021}.
\citeauthor{cinelli2020covid19} (\citeyear{cinelli2020covid19}) studied rumor amplification in five social media platforms, where rumors were labeled using distant supervision.
In contrast, we have careful manual annotation and multiple labels. \citeauthor{zhou2020repository} created the ReCOVery dataset, which combines news articles about COVID-19 with tweets. \citeauthor{vidgen2020detecting} (\citeyear{vidgen2020detecting}) studied COVID-19 prejudices using a manually labeled dataset of 20K tweets with the following labels: hostile, criticism, prejudice, and neutral.
The closest work to ours is that of \citeauthor{song2020classification} (\citeyear{song2020classification}), who collected a dataset of false and misleading claims about COVID-19 from IFCN Poynter, which they manually annotated with ten disinformation categories:
(1)~Public authority,
(2)~Community spread and impact, 
(3)~Medical advice, self-treatments, and virus effects,
(4)~Prominent actors,
(5)~Conspiracies,
(6)~Virus transmission,
(7)~Virus origins and properties,
(8)~Public reaction, and
(9)~Vaccines, medical treatments, and tests,
and
(10)~Cannot determine.
Their categories partially overlap with ours, but ours are broader and account for more perspectives. Moreover, we cover both true and false claims, we focus on tweets (while they have general claims), and we cover both English and Arabic (they only cover English).
Other related work is FakeCovid \cite{shahi2020fakecovid}, a multilingual cross-domain dataset consisting of manually labeled 1,951 articles. The study by \cite{pulido2020covid} analyzed 1,000 tweets and categorized them based on factuality: {\em (i)} False information, {\em (ii)} Science-based evidence, {\em (iii)} Fact-checking tweets, {\em (iv)} Mixed information, {\em (v)} Facts, {\em (vi)} Facts, {\em (vii)} Other, and {\em (viii)} Not valid. 
Finally, \cite{ding2020challenges} have a position paper discussing the challenges in combating the COVID-19 infodemic in terms of data, tools, and ethics. 
See also a recent survey: \cite{Shuja2020.05.19.20107532}.

%% file: sections/annotation_platform.tex
\section{Annotation Setup}
\label{sec:annotation_platform}

Below, we present the annotation schema that we developed after a lot of analysis and discussion, and which we refined during the pilot annotations. We then present the annotation platform and interface we used. 

\begin{figure*}[htb!]
\centering
\includegraphics[width=0.7\textwidth]{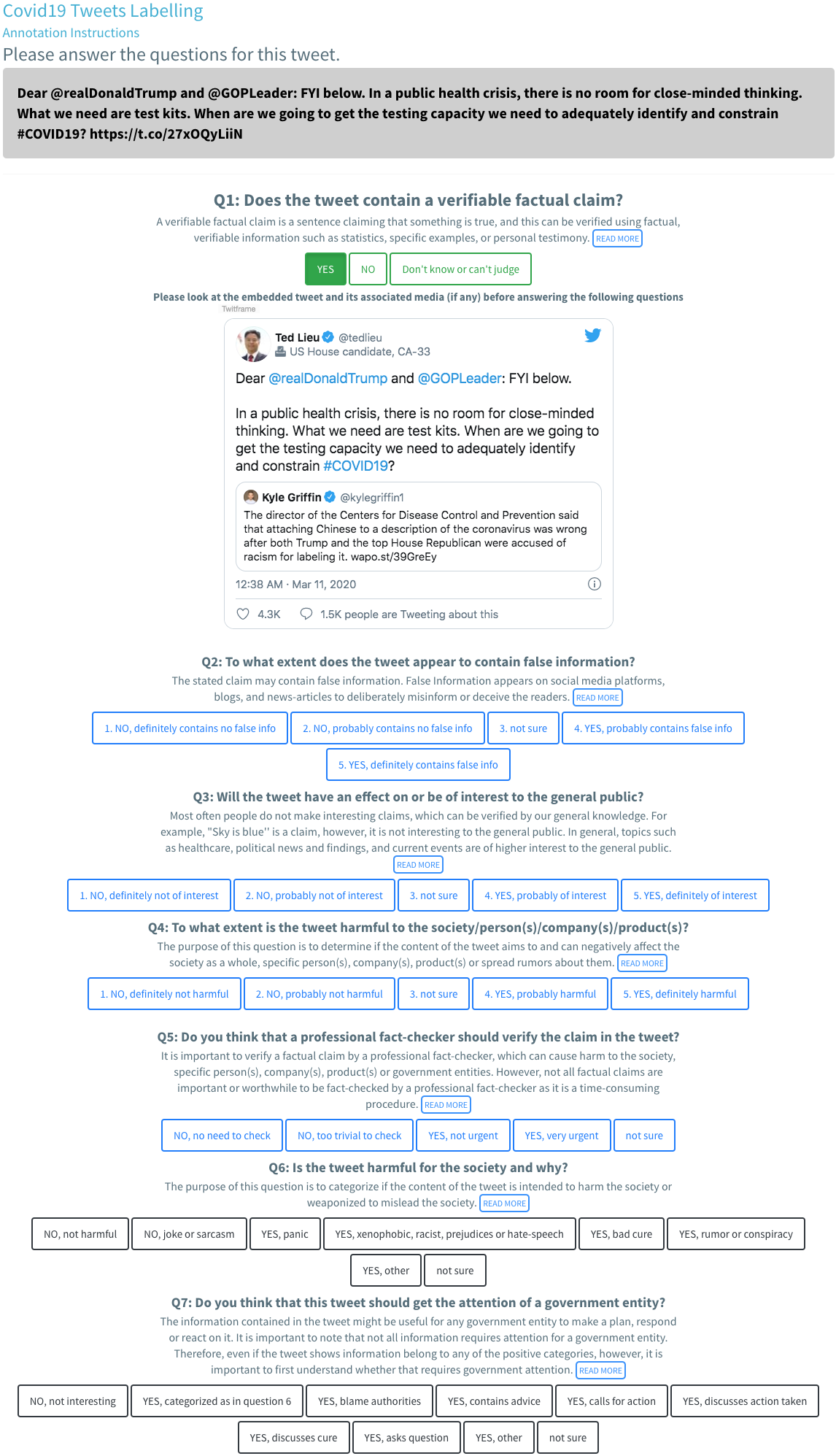}
\caption{\textbf{The platform for an English tweet:} a \emph{Yes} answer for Q1 has shown questions Q2--Q7 and their answers.}
\label{fig:example_english_tweet}
\end{figure*}

\begin{figure*}[htb!]
\centering
\includegraphics[width=0.65\textwidth]{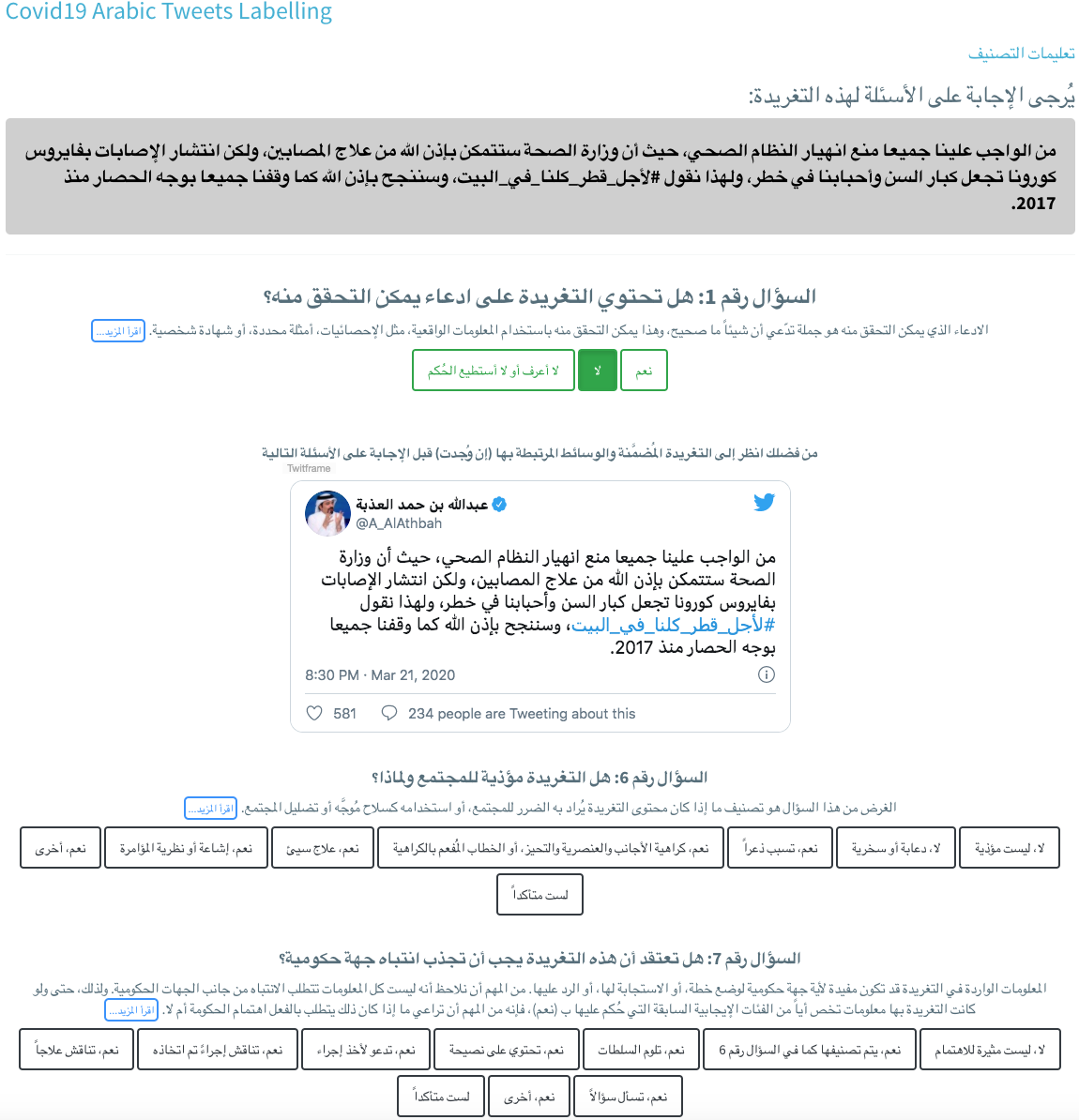}
\caption{\textbf{The platform for an Arabic tweet:} a \emph{No} answer for Q1 means that only Q6 and Q7 would be shown. (English translation of the Arabic text in the tweet: \emph{We must prevent the collapse of the healthcare system. The Ministry of Public Health will cure the infected people, but the spread of the infection puts the elderly and our beloved ones in danger. That is why we say \#StayHomeForQatar, and we will succeed...})}
\label{fig:example_arabic_tweet}
\end{figure*}

\subsection{Annotation Schema and Instructions}

We designed the annotation instructions after careful analysis and discussion, followed by iterative refinement based on observations from the pilot annotation. Our annotation schema is organized into seven questions.

\input{sections/annotation_questions}

A notable property of our schema is that the fine-grained labels can be easily transformed into coarse-grained binary YES/NO labels, i.e.,~all no* labels could be merged into a \emph{NO} label, and all yes* labels can become \emph{YES}. Note also that some questions (i.e., Q2, Q3, and Q4) are on an ordinal scale, and thus can be addressed using ordinal regression.

Finally, note that even though our annotation instructions were developed to analyze the COVID-19 infodemic, they can be potentially adapted for other kinds of global crises, where taking multiple perspectives into account is desirable.

\subsection{Annotation Platform}
\label{ssec:annotation_platform}

Our crowd-sourcing annotation platform is based on MicroMappers,\textsuperscript{\ref{micromapper}}
a framework that was used for several disaster-related social media volunteer annotation campaigns in the past. We configured MicroMappers to allow labeling COVID-19 tweets in English and Arabic for all seven questions.
Initially, the interface only shows the text of the tweet and the answer options for Q1.
Then, depending on the selected answer, it dynamically shows either Q2-Q7 or Q6-Q7. After Q1 has been answered, it shows not just the text of the tweet, but its actual look and feel as it appears on Twitter. The annotation instructions are quickly accessible at any moment for the annotators to check. 

Figure \ref{fig:example_english_tweet} shows an example of an English tweet, where the answer \emph{Yes} was selected for Q1, which has resulted in displaying the tweet as it would appear in Twitter as well as showing all the remaining questions with their associated answers. Figure~\ref{fig:example_arabic_tweet} shows an Arabic example, where a \emph{No} answer was selected,\footnote{Note that this answer is actually wrong, as there are verifiable factual claims in the tweet. Here, it was selected for demonstration purposes only.} which has resulted in showing questions Q6 and Q7 only.

Using the annotation platform has reduced our in-house annotation efforts significantly, cutting the annotation time by half compared to using a spreadsheet, and we expect similar time savings for general crowd-sourcing annotations. The platform is collaborative in nature, and multiple annotators can work on it simultaneously. In order to ensure the quality of the annotations, we have configured the platform to require five annotators per tweet.

%% file: sections/annotation_questions.tex
\subsubsection{Q1: Does the tweet contain a verifiable factual claim?}
This is an objective question, and it proved very easy to annotate. Positive examples include\footnote{This is influenced by \cite{DBLP:journals/corr/abs-1809-08193}.} tweets that state a definition, mention a quantity in the present or in the past, make a verifiable prediction about the future, reference laws, procedures, and rules of operation, discuss images or videos, and state correlation or causation, among others. 

We show the annotator the tweet text only, and we ask her to answer the question, without checking anything else. 
This is a \emph{Yes}/\emph{No} question, but we also have a \emph{Don't know or can't judge} answer, which is to be used in tricky cases, e.g.,~when the tweet is not in English or Arabic. 
If the annotator selects \emph{Yes}, then questions 2--5 are to be answered as well; otherwise, they are skipped automatically (see Section \ref{ssec:annotation_platform}).

\subsubsection{Q2: To what extent does the tweet appear to contain false information?}
This question asks for a subjective judgment; it does not ask for annotating the actual factuality of the claim in the tweet, but rather whether the claim \emph{appears} to be false.
For this question (and for all subsequent questions), we show the tweet as it is displayed in the Twitter feed, which can reveal some useful additional information, e.g.,~a link to an article from a reputable information source could make the annotator more likely to believe that the claim is true. The annotation is on a 5-point ordinal scale:

\begin{enumerate}
\itemsep0em 
\item \emph{NO, definitely contains no false information}
\item \emph{NO, probably contains no false information}
\item \emph{not sure}
\item \emph{YES, probably contains false information}
\item \emph{YES, definitely contains false information}
\end{enumerate}

\subsubsection{Q3: Will the tweet have an effect on or be of interest to the general public?}
Generally, claims that contain information related to potential cures, updates on number of cases, on measures taken by governments, or discussing rumors and spreading conspiracy theories should be of general public interest. Similarly to Q2, the labels are defined on a 5-point ordinal scale; however, unlike Q2, this question is partially objective (the \emph{YES}/\emph{NO} part) and partially subjective (the \emph{definitely}/\emph{probably} distinction).

\begin{enumerate}
\itemsep0em 
\item \emph{NO, definitely not of interest}
\item \emph{NO, probably not of interest}
\item \emph{not sure}
\item \emph{YES, probably of interest}
\item \emph{YES, definitely of interest}
\end{enumerate}

\subsubsection{Q4: To what extent is the tweet harmful to the society,  person(s), company(s) or product(s)?}
This question asks to identify tweets that can negatively affect society as a whole, but also specific person(s), company(s), product(s). The labels are again on a 5-point ordinal scale, and, similarly to Q3, this question is partially objective (\emph{YES}/\emph{NO}) and partially subjective (\emph{definitely}/\emph{probably}).

\begin{enumerate}
\itemsep0em 
\item \emph{NO, definitely not harmful}
\item \emph{NO, probably not harmful}
\item \emph{not sure}
\item \emph{YES, probably harmful}
\item \emph{YES, definitely harmful}
\end{enumerate}

\subsubsection{Q5: Do you think that a professional fact-checker should verify the claim in the tweet?}
This question asks for a subjective opinion. Yet, its answer should be informed by the answer to questions Q2, Q3 and Q4, as a check-worthy factual claim is probably one that is likely to be false, is of public interest, and/or appears to be harmful. Here the answers are not on an ordinal scale, but rather focus on the reason why there is or is not a need to fact-check the tweet:

\begin{enumerate}[label=\Alph*.]
\itemsep0em 
\item \emph{NO, no need to check}: there is no need to fact-check the claims(s) made in the tweet, e.g.,~because they are not interesting, make a joke, etc.
\item \emph{NO, too trivial to check}: the tweet is worth fact-checking, but this does not require a professional fact-checker, i.e.,~a non-expert might be able to fact-check it easily, e.g.,~by using reliable sources such as the official website of the World Health Organization, Wikipedia, etc. An example of such a claim is as follows: ``\emph{China has 24 times more people than Italy...}''
\item \emph{YES, not urgent}: the tweet should be fact-checked by a professional fact-checker, but this is not urgent, nor is it critical.
\item \emph{YES, very urgent}: the tweet can cause immediate harm to a large number of people, and thus it should be fact-checked as soon as possible by a professional fact-checker.
\item \emph{not sure}: the tweet does not contain enough information to allow for a clear judgment on whether it is worth fact-checking, or the annotator is simply not sure. 
\end{enumerate}

\subsubsection{Q6: Is the tweet harmful to the society and why?}
\label{ssec:question_task_6}

This is an objective question.
It asks whether the tweet is harmful to the society (unlike Q4, which covers broader harm, e.g.,~to persons, companies, and products). It further asks to categorize the nature of the harm, if any. Similarly to Q5 (and unlike Q4), the answers are categorical and are not on an ordinal scale.

\begin{enumerate}[label=\Alph*.]
\itemsep0em 
\item \emph{NO, not harmful}: the tweet is not harmful to the society.
\item \emph{NO, joke or sarcasm}: the tweet contains a joke or expresses sarcasm.
\item \emph{not sure}: the content of the tweet makes it hard to make a judgment.
\item \emph{YES, panic}: the tweet can cause panic, fear, or anxiety.
\item \emph{YES, xenophobic, racist, prejudices, or hate-speech}: the tweet contains a statement that relates to xenophobia, racism, prejudices, or hate speech.
\item \emph{YES, bad cure}: the tweet promotes a questionable cure, medicine, vaccine, or prevention procedures.
\item \emph{YES, rumor, or conspiracy}: the tweet spreads rumors or conspiracy theories.
\item \emph{YES, other}: the tweet is harmful, but it does not belong to any of the above categories.
\end{enumerate}

\subsubsection{Q7: Do you think that this tweet should get the attention of a government entity?}

This question asks for a subjective judgment (unlike Q6 which was objective) about whether the target tweet should get the attention of a government entity or of policy makers in general. Similarly to Q5 and Q6, the answers to this question are categorical and are not on an ordinal scale.

\begin{enumerate}[label=\Alph*.]
\itemsep0em 
\item \emph{NO, not interesting}: the tweet is not interesting for any government entity.
\item \emph{not sure}: the content of the tweet makes it hard to make a judgment.
\item \emph{YES, categorized as in Q6}:
a government entity should pay attention to this tweet as it was labeled with some of the \textit{YES} sub-categories in Q6.
\item \emph{YES, other}: the tweet needs the attention of a government entity, but it cannot be labeled as any of the above categories.
\item \emph{YES, blames authorities}: the tweet blames government authorities or top politicians.
\item \emph{YES, contains advice}: the tweet contains advice about some COVID-19 related social, political, national, or international issues that might be of interest to a government entity.
\item \emph{YES, calls for action}: the tweet states that some government entities should take action on a particular issue.
\item \emph{YES, discusses action taken}: the tweet discusses specific actions or measures taken by governments, companies, or individuals regarding COVID-19.
\item \emph{YES, discusses cure}: the tweet discusses possible cure, vaccine or treatment for COVID-19.
\item \emph{YES, asks a question}: 
the tweet raises question that might need an official answer.
\end{enumerate}

%% file: sections/dataset.tex
\section{Pilot Annotation Dataset}
\label{dataset}

\subsection{Data for the Pilot Annotation}
\label{sec:data:pilot}

We collected frequent tweets (i.e., such with at least 500 retweets) about COVID-19 in March 2020, in both English and Arabic. We used twarc\footnote{\url{http://github.com/DocNow/twarc}} for crawling. To collect the tweets, we used the following keywords and hashtags for English: 

\begin{itemize}
	\item \emph{\#covid19, \#CoronavirusOutbreak, \#Coronavirus, \#Corona, \#CoronaAlert, \#CoronaOutbreak, Corona, covid-19}. 
\end{itemize}

For Arabic, we used corresponding Arabic equivalents.

\subsection{Annotation}

We performed a pilot annotation in order to test the platform and to refine the annotation guidelines. We annotated 504 English tweets for questions Q1, Q6, and Q7; however, we have 305 tweets for questions Q2, Q3, Q4, and Q5 as they are only annotated if the answer to Q1 is \emph{Yes}. Similarly, for Arabic, we have 218 tweets for Q1, Q6, and Q7, and 140 tweets for Q2, Q3, Q4, and Q5.

We performed the annotation in three stages. In the first stage, 2--5 annotators independently annotated a batch of 25-50 examples. In the second stage, these annotators met to discuss and to try to resolve the cases of disagreement. In the third stage, any unresolved cases were discussed in a meeting involving all authors of this paper. 

In stages two and three, we further discussed whether handling the problematic tweets required adjustments or clarifications in the annotation guidelines. In case of any such change for some questions, we reconsidered all previous annotations for that question in order to make sure the annotations reflected the latest version of the annotation guidelines.

\begin{table}[t!]
\centering
\scalebox{0.60}{
\begin{tabular}{llrr}
\toprule
\multicolumn{1}{c}{\textbf{Exp.}} & \multicolumn{1}{c}{\textbf{Class labels}} & \multicolumn{1}{c}{\textbf{EN}} & \multicolumn{1}{c}{\textbf{AR}} \\ \midrule
\multicolumn{2}{l}{\textbf{\begin{tabular}[c]{@{}l@{}}Q1: Does the tweet contain a verifiable factual claim?\end{tabular}}} & \textbf{504} & \textbf{218} \\ \midrule
 & No & 199 & 78 \\
\multirow{-2}{*}{Bin} & Yes & 305 & 140 \\
\rowcolor[HTML]{EBF5FB} \midrule
\multicolumn{2}{l}{\textbf{\begin{tabular}[c]{@{}l@{}}Q2: To what extent does the tweet appear \\  to contain false information?\end{tabular}}} & \textbf{305} & \textbf{140} \\ \midrule
\rowcolor[HTML]{EBF5FB}
 & No, definitely contains no false info & 46 & 31 \\
\rowcolor[HTML]{EBF5FB}
 & No, probably contains no false   info & 177 & 62 \\
\rowcolor[HTML]{EBF5FB}
 & not sure & 45 & 5 \\
\rowcolor[HTML]{EBF5FB}
 & Yes, probably contains false   info & 25 & 40 \\
\rowcolor[HTML]{EBF5FB}
\multirow{-5}{*}{Multi} & Yes, definitely contains false   info & 12 & 2 \\
\rowcolor[HTML]{EBF5FB} \midrule
 & No & 223 & 93 \\
\rowcolor[HTML]{EBF5FB}
\multirow{-2}{*}{Bin} & Yes & 37 & 42 \\ \midrule
\multicolumn{2}{l}{\textbf{\begin{tabular}[c]{@{}l@{}}Q3: Will the tweet's   claim have an effect on or  \\ be of interest to the general public?\end{tabular}}} & \textbf{305} & \textbf{140} \\ \midrule
 & No, definitely not of interest & 10 & 1 \\
 & No, probably not of interest & 46 & 5 \\
 & not sure & 8 & 9 \\
 & Yes, probably of interest & 180 & 76 \\
\multirow{-5}{*}{Multi} & Yes, definitely of interest & 61 & 49 \\ \midrule
 & No & 56 & 6 \\
\multirow{-2}{*}{Bin} & Yes & 241 & 125 \\ \midrule
\rowcolor[HTML]{EBF5FB}
\multicolumn{2}{l}{\textbf{\begin{tabular}[c]{@{}l@{}}Q4: To what extent   does the tweet appear to \\ be harmful to society, person(s), company(s) or   product(s)?\end{tabular}}} & \textbf{305} & \textbf{140} \\ \midrule
\rowcolor[HTML]{EBF5FB}
 & No, definitely not harmful & 111 & 68 \\
\rowcolor[HTML]{EBF5FB}
 & No, probably not harmful & 67 & 21 \\
\rowcolor[HTML]{EBF5FB}
 & not, sure & 2 & 3 \\
\rowcolor[HTML]{EBF5FB}
 & Yes, probably harmful & 67 & 46 \\
\rowcolor[HTML]{EBF5FB}
\multirow{-5}{*}{Multi} & Yes, definitely harmful & 58 & 2 \\ \midrule
\rowcolor[HTML]{EBF5FB}
 & No & 178 & 89 \\
\rowcolor[HTML]{EBF5FB}
\multirow{-2}{*}{Bin} & Yes & 125 & 48 \\ \midrule
\multicolumn{2}{l}{\textbf{\begin{tabular}[c]{@{}l@{}}Q5: Do you think   that a professional fact-checker \\ should verify the claim in the tweet?\end{tabular}}} & \textbf{305} & \textbf{140} \\  \midrule
 & No, no need to check & 81 & 22 \\
 & No, too trivial to check & 64 & 55 \\
 & Yes, not urgent & 117 & 48 \\
 \multirow{-4}{*}{Multi} & Yes, very urgent & 43 & 15 \\  \midrule
 & No & 145 & 77 \\
\multirow{-2}{*}{Bin} & Yes & 160 & 63 \\  \midrule
\rowcolor[HTML]{EBF5FB}
\multicolumn{2}{l}{\textbf{Q6:   Is the tweet harmful for society and why?}} & \textbf{504} & \textbf{218} \\ \midrule
\rowcolor[HTML]{EBF5FB}
 & No, joke or sarcasm & 62 & 2 \\
\rowcolor[HTML]{EBF5FB}
 & No, not harmful & 333 & 159 \\
\rowcolor[HTML]{EBF5FB}
 & not sure & 2 & 0 \\
\rowcolor[HTML]{EBF5FB}
 & Yes, bad cure & 3 & 1 \\
\rowcolor[HTML]{EBF5FB}
 & Yes, other & 25 & 5 \\
\rowcolor[HTML]{EBF5FB}
 & Yes, panic & 23 & 12 \\
\rowcolor[HTML]{EBF5FB}
 & Yes, rumor conspiracy & 42 & 33 \\
\rowcolor[HTML]{EBF5FB}
\multirow{-8}{*}{Multi} & \begin{tabular}[c]{@{}l@{}}Yes, xenophobic racist prejudices  or hate   speech\end{tabular} & 14 & 6 \\ \midrule
\rowcolor[HTML]{EBF5FB}
 & No & 395 & 161 \\
\rowcolor[HTML]{EBF5FB}
\multirow{-2}{*}{Bin} & Yes & 107 & 57 \\ \midrule
\multicolumn{2}{l}{\textbf{\begin{tabular}[c]{@{}l@{}}Q7: Do you think   that this tweet should \\get the attention of any government entity?\end{tabular}}} & \textbf{504} & \textbf{218} \\ \midrule
 & No, not interesting & 319 & 163 \\
 & not sure & 6 & 0 \\
 & Yes, asks question & 2 & 0 \\
 & Yes, blame authorities & 81 & 13 \\
 & Yes, calls for action & 8 & 1 \\
 & Yes, classified as in question   6 & 34 & 30 \\
 & Yes, contains advice & 9 & 1 \\
 & Yes, discusses action taken & 12 & 6 \\
 & Yes, discusses cure & 5 & 4 \\ 
\multirow{-10}{*}{Multi} & Yes, other & 28 & 0 \\ \midrule
 & No & 319 & 163 \\
\multirow{-2}{*}{Bin} & Yes & 179 & 55 \\ 
\bottomrule
\end{tabular}
}
\caption{Distribution for the English and the Arabic datasets. In the rows with a question, we show the total number of tweets for the respective language. For the binary task (Bin), we map all multiclass (Multi) \textit{Yes*} labels to \textbf{Yes}, and the \textit{No*} labels to \textbf{No}, and we further drop the \textit{not sure} labels.}
\label{tab:class_label_distribution}
\end{table}

\subsection{Annotation Agreement}
In the process of annotation, we were calculating the current inter-annotator agreement. Fleiss Kappa was generally high for objective questions, e.g., it was over 0.9 for Q1, and around 0.5 for Q6. For subjective and partially subjective questions, the scores ranged around 0.4 and 0.5, with the notable exception of Q5 with 0.8. Note that values of Kappa of 0.21--0.40, 0.41--0.60, 0.61--0.80, and 0.81--1.0 correspond to fair, moderate, substantial and perfect agreement, respectively~\cite{landis1977measurement}.

\subsection{Data Statistics}
Table \ref{tab:class_label_distribution} shows statistics about the annotations. While we focus the following analysis on English tweets, the distribution of the Arabic ones is similar. 

The class distribution for Q1 is quite balanced, (61\% YES and 39\% NO examples). Recall that only the tweets that are labeled as factual were annotated for Q2-5.  For question Q2, the label ``No, probably contains no false info'' is frequent, which means that most tweets considered credible. Out of 305 tweets labeled for Q2, about 73\% are judged to contain no false information, whereas 12\% were categorized as ``not sure'', and 15\% as ``contains false information'', either ``probably'' or ``definitely''.
  
For Q3, which asks whether \emph{the tweet is of interest to the general public}, the distribution is skewed towards \emph{Yes} in 79\% of the cases. This can be attributed to the tweets having been selected based on frequency of retweets and likes.

For Q4, which asks whether \emph{the tweet is harmful to the society}, the labels for the tweets vary from not harmful to harmful, covering all cases, without huge spikes. 

For Q5, which asks whether \emph{a professional fact-checkers should verify the claim}, the majority of the cases were labeled as either ``Yes, not urgent'' (38\%) or ``No, no need to check'' (27\%). It appears that a professional fact-checker should verify the claims made in the tweets immediately in only a small number of cases (14\%). 

For questions Q3-5, the ``not sure'' cases were generally very few. Yet, such cases were substantially more prevalent in the case of Q2. Identifying potentially false claims (Q2) is challenging, as it might require external information. When annotating Q2, the annotators were shown the entire tweet, and they could further open tweet and see the entire thread in Twitter,

For Q6, most of the tweets were classified as ``not harmful'' for the society or as a ``joke or sarcasm''. From the critical classes, 3\% of the tweets are classified as containing ``xenophobic, racist, prejudices or hate speech'', and 5\% as ``spreading panic''. 

For Q7, it is clear that, in the majority of cases (64\%), the tweets are not of interest to government entities and policy makers; yet, 16\% of the tweets blame the authorities. 

%% file: sections/experiments_results.tex
\section{Experiments and Evaluation}
\label{sec:experiments_results}

\begin{table*}[tbh]
\footnotesize
\centering
\scalebox{0.73}{
\begin{tabular}{@{}c@{}c@{}c@{}c@{}c@{}c@{}c@{}c@{}|@{}c@{}c@{}c@{}c@{}c@{}}
\toprule
 \multicolumn{8}{c|}{\textbf{English}} & \multicolumn{5}{c}{\textbf{Arabic}} \\ \cmidrule{3-8} \cmidrule{9-13}
\multicolumn{1}{l}{\textbf{Q.}} & \multicolumn{1}{l}{\textbf{Cls}} & \multicolumn{1}{c}{\textbf{Maj.}} & \multicolumn{1}{c}{\textbf{FastText}}& \multicolumn{1}{c}{\textbf{BERT}} & \multicolumn{1}{c}{\textbf{mBERT}} & \multicolumn{1}{c}{\textbf{RoBERTa}} & \multicolumn{1}{c|}{\textbf{ALBERT }} & \multicolumn{1}{c}{\textbf{Maj.}} & \multicolumn{1}{c}{\textbf{FastText}}& \multicolumn{1}{c}{\textbf{mBERT}} & \multicolumn{1}{c}{\textbf{AraBERT}} & \multicolumn{1}{c}{\textbf{XLM-r}} \\ \cmidrule{3-8} \cmidrule{9-13}
\multicolumn{13}{c}{\textbf{Binary (Coarse-grained)}} \\\midrule
Q1 & 2 & 45.6 & \textbf{72.8} & \bf 87.6& \textbf{88.3} & \bf \underline{90.6} & \bf 86.5 & 50.2 & \textbf{75.8} & \bf \underline{88.1} & \bf 82.6 & \bf 76.9 \\
Q2 & 2 & 79.2 & \textbf{82.6} & \bf \underline{86.9}& \textbf{83.1} & \bf 82.9 & \bf 83.9 & 56.2 & \bf 68.2 &  \bf \underline{79.1} & \bf 71.1 & \bf 60.2 \\
Q3 & 2 & 72.7 & \textbf{77.2} & \bf \underline{84.3}& \textbf{81.6} & \bf 80.8 & \bf 79.6 &  \underline{93.2} & \underline{93.2} & 89.2 & 77.8 & 89.2 \\
Q4 & 2 & 43.5 & \textbf{69.6} & \bf \underline{84.0}& \textbf{82.7} & \bf 83.8 & \bf 78.5 & 51.2 & \bf 79.2 & \bf 78.5 & \bf \underline{80.4} & \bf 69.0 \\
Q5 & 2 & 36.1 & \textbf{63.1} & \bf \underline{81.3}& \textbf{80.0} & \bf 73.7 & \bf 72.7 & 39.0 & \bf \underline{78.6} &  \bf 76.4 & \bf 76.1 & \bf 66.5 \\
Q6 & 2 & 69.3 & \textbf{71.6} & \bf \underline{86.1}& \textbf{76.8} & \bf 81.0 & \bf 79.2 & 62.7 & \bf 79.4 & \bf \underline{80.4} & \bf 77.3 & \bf 64.6 \\
Q7 & 2 & 50.0 & \textbf{69.9} & \bf \underline{89.3}& \textbf{81.9} & \bf \bf 84.7 & \bf 79.0 & 64.0 & \bf \bf 74.1 & \bf \underline{78.5} & \bf 77.9 & 64.0 \\ \midrule
\multicolumn{13}{c}{\textbf{Multiclass (Fine-grained)}} \\\midrule
Q2 & 5 & 42.6 & \textbf{44.0} &  \bf 48.5 & \bf \underline{52.2} & \bf 46.6 & \bf 44.8 & 27.2 & \bf \underline{47.4} &  \bf 42.8 & \bf 42.1 & \bf 37.4 \\
Q3 & 5 & 43.8 & \textbf{48.3} &  \bf \underline{57.6}& \textbf{45.1} & \bf 50.9 & \bf 45.4 &  \bf 38.2 & \bf \underline{83.1} & 27.0 & 21.4 & 20.0 \\
Q4 & 5 & 19.4 & \textbf{35.5} &  \bf 41.6 & \textbf{42.9} & \bf \underline{44.1} & \bf 39.5 & 31.8 & \bf \underline{54.4} &  \bf 43.7 & \bf 44.9 & \bf 34.2 \\
Q5 & 5 & 21.3 & \textbf{37.6} &  \bf 50.4& \bf \underline{52.3} & \bf 50.3 & \bf 48.0 & 22.2 & \bf \underline{77.2} &  \bf 59.0 & \bf 57.7 & \bf 46.1 \\
Q6 & 8$^{*}$ & 52.6 & \textbf{53.9} &  \bf 57.2 & \bf \underline{62.7} & \bf 58.4 & \bf 56.5 & 61.5 & \bf \underline{79.3} & 40.9 & 38.9 & 44.5 \\
Q7 & 10$^{*}$ & 49.1 & \textbf{57.8} &  \bf 54.6 & \bf \underline{58.7} & \bf 55.2 & \bf 53.5 & 64.0 & \bf \underline{75.7} &  \bf 66.3 & 63.9 & 64.0 \\
\bottomrule
\end{tabular}
}
\caption{\textbf{Experiments using different models.} Binary and multiclass results (weighted F1), for English and Arabic, using various Transformers and FastText. The results that improve over the majority class baseline (\emph{Maj.}) are in \textbf{bold}, and the best system is \underline{underlined}. Legend: Q. -- question, Cls -- number of classes, the * in Q6 and Q7 is a reminder that for Arabic there are 7 classes (not 8 and 10 as for English). }
\label{tab:results_binary_multiclass_tasks}
\vspace{-1.2em}
\end{table*}

\subsection{Experimental Setup}

We performed experiments using both binary and multiclass settings. We first performed standard pre-processing of the tweets: removing hash tags and other symbols, and replacing URLs and usernames by special tags. 
Due to the small size of the datasets, we used 10-fold cross validation. To tune the hyper-parameters of the models, we split each training fold into \texttt{train}$_{train}$ and \texttt{train}$_{dev}$ parts, and we used the latter for finding the best hyper-parameter values.

\subsubsection{Models} Large-scale pre-trained transformers have achieved state-of-the-art performance for several NLP tasks. We experimented with such models and binary vs. multiclass, low-resource task scenarios.
More specifically, we used BERT \citep{devlin2018bert}, RoBERTa \citep{liu2019roberta}, and ALBERT \citep{lan2019albert} for English, and multilingual BERT (mBERT), XLM-r~\citep{conneau2019unsupervised} and AraBERT \citep{baly2020arabert} for Arabic. In addition to pre-trained models, we also evaluated the performance of static-embedding based classification using FastText~\citep{joulin2017bag}. 

For transformer-based models, 
we fine-tuned each model using the default settings for three epochs as described in \cite{devlin2018bert}. Due to instability, we performed ten runs of each experiment using different random seeds, and we picked the model that performs the best on the development set. For FastText, we used embeddings trained on Common Crawl.

\subsubsection{Evaluation Measure} We report weighted F1 score, as it takes class imbalance into account.

\subsection{Results}
\label{sec:evaluation}

\subsubsection{Baseline}

We use a simple majority class baseline. Note that for questions with highly imbalanced label distribution, it can achieve very high scores. For example, for Q3 in the binary setting for Arabic, 125 out of 131 tweets are in the `Yes' category, which yields a balanced F1 score of 93\% for the majority class baseline.

\subsubsection{Binary Classification}
The first part of Table \ref{tab:results_binary_multiclass_tasks} presents the results for binary classification using various models. 

\textbf{Results for English:}
We can see that all models performed better than the majority class baseline and FastText, confirming the efficacy of transformers.
Comparing various pre-trained models, we can see that BERT outperformed all other models on six out of the seven tasks, while ALBERT performed the worst in most of the cases. For Q1, RoBERTa and mBERT performed better than BERT, with RoBERTa performing the best.

\textbf{Results for Arabic:}
For all tasks except for Q3 Arabic (which has very skewed distribution), the models performed better than the majority class baseline. 
Unlike English, this time, there was no model that outperformed the rest overall. We can see that XLM-r performed worse, that mBERT outperformed all the other models for four out of seven tasks, and that AraBERT performed better than other models for Q4. Interestingly, FastText performed very well on many tasks, achieving the best overall results on Q5. This could be due to it using character n-grams, which can be important for a morphologically rich language such as Arabic.

\subsubsection{Multiclass Classification}
The second part of Table \ref{tab:results_binary_multiclass_tasks} shows the results in the multiclass setting. The \emph{Cls} column shows the number of classes per task, and we can see that the number of classes now increases from 2 to 5--10, depending on the question. This makes the classification tasks much harder, which is reflected in the substantially lower weighted F1 scores, both for the baselines and for the models we experimented with.

\textbf{Results for English:}
We can see that all models performed better than the majority class. The most successful one was mBERT, which performed the best in four out of six tasks. Interestingly, mBERT outperformed BERT in several cases.

\textbf{Results for Arabic:}
This time, FastText outperformed all transformers models. Once again, this can be due to it using embeddings for character n-grams, which makes it more robust to morhphological variations in the input, including possible typos. This could also indicate the training data not being sufficient to optimize the large number of parameters in the transformer models.

%% file: sections/conclusion.tex
\section{Conclusion and Future Work}
\label{sec:conclutions}

In a bid to effectively counter the first global infodemic related to COVID-19, we have argued for the need for a holistic approach combining the perspectives of journalists, fact-checkers, policymakers, government entities, social media platforms, and society.
With this in mind and in order to reduce the annotation effort and to increase the quality of the annotations, we have developed a volunteer-based crowd annotation tools based on the MicroMappers platform. Now, we issue a \emph{call to arms} to the research community and beyond to join the fight by supporting our crowd-sourcing annotation efforts.
We plan to support the annotation platforms with fresh tweets. We further plan to release annotation platforms for other languages. Finally, we plan regular releases of the data obtained thanks to the crowdsourcing efforts.

%% file: sections/supplemental_material.tex


\section{Detail Annotation Instructions}
\label{sec:annotation_instruc_detail}

\paragraph{General Instructions:}
\begin{enumerate}
     \itemsep-0.2em 
    \item For each tweet, the annotator needs to read the text including the hashtags and also look at the tweet itself when necessary by going to the link (i.e., for Q2-7 it might be required to open the tweet link).\footnote{The reason for not going to the tweet link for Q1 is that we wanted to reduce the complexity of the annotation task and to focus on the content of the tweet only. As for Q2, it might be important to check if the tweet was posted by an authoritative source, and thus it might be useful for the annotator to open the tweet to get more context. After all, this is how real users perceive the tweet. Since the annotators would open the tweet's link for Q2, they can use that information for the rest of the questions as well (even though this is not required). 
    \item The annotators should assume the time when the tweet was posted as a reference when making judgments, e.g., \textit{``Trump thinks, that for the vast majority of Americans, the risk is very, very low.''} would be true when he made the statement but false by the time annotations were carried out for this tweet. The annotator should consider the time when the tweet was posted.} 
    \item The annotators may look at the images and the videos, to the Web pages that the tweet links to, as well as to the tweets in the same thread when making a judgment, if required.
    \item The annotators are not required to complete questions Q2-Q5 if the answer to question Q1 is \lbl{NO}.
\end{enumerate}


\subsection{Verifiable Factual Claim} 
\uline {\textbf{Question 1:} Does the tweet contain a verifiable factual claim?}

A \emph{verifiable factual claim} is a sentence claiming that something is true, and this can be verified using factual verifiable information such as statistics, specific examples, or personal testimony. 
Factual claims include the following:\footnote{Inspired by \citep{DBLP:journals/corr/abs-1809-08193}.}
\begin{itemize}
    \itemsep-0.2em 
    \item Stating a definition; 
    \item Mentioning quantity in the present or the past;
    \item Making a verifiable prediction about the future; 
    \item Statistics or specific examples;
    \item Personal experience or statement (e.g., \textit{``I spent much of the last decade working to develop an \#Ebola treatment.''})
    \item Reference to laws, procedures, and rules of operation;
    \item References (e.g., URL) to images or videos (e.g., \textit{``This is a video showing a hospital in Spain.''});
    \item Statements which can be technically classified as questions, but in fact contain a verifiable claim based on the criteria above (e.g., \textit{``Hold on - \#China Communist Party now denying \#CoronavirusOutbreak originated in China? This after Beijing's catastrophic mishandling of the virus has caused a global health crisis?''})
    \item Statements about correlation or causation. Such a correlation or causation needs to be explicit, i.e., sentences like \textit{``This is why the beaches haven't closed in Florida. https://t.co/8x2tcQeg21''} is not a claim because it does not explicitly say why, thus it is not verifiable. 
\end{itemize}
Tweets containing personal opinions and preferences are not factual claims. 
Note that if a tweet is composed of multiple sentences or clauses, at least one full sentence or clause needs to be a claim in order for the tweet to contain a factual claim. If a claim exist in a sub-sentence or sub-clause, then the tweet is not considered to have a factual claim. 
For example, \textit{``My new favorite thing is Italian mayors and regional presidents LOSING IT at people violating quarantine''} is not a claim -- it is in fact an opinion. Moreover, if we consider \textit{``Italian mayors and regional presidents LOSING IT at people violating quarantine''} it would be a claim. In addition, when answering this question, annotators should not open the tweet URL. 
Since this is a binary decision task, the answer of this question consists of two labels as defined below. 

\noindent\textbf{Labels:}
\begin{itemize}
    \itemsep-0.2em 
    \item \lbl{YES:} if it contains a verifiable factual claim;
    \item \lbl{NO:} if it does not contain a verifiable factual claim;
    \item \lbl{Don't know or can't judge:} the content of the tweet does not have enough information to make a judgment. It is recommended to categorize the tweet using this label when the content of the tweet is not understandable at all. For example, it uses a language (i.e., non-English) or references that it is difficult to understand;
\end{itemize}

\noindent \textbf{Examples:}

\begin{enumerate}
    \item \textit{Please don't take hydroxychloroquine (Plaquenil) plus Azithromycin for \#COVID19 UNLESS your doctor prescribes it. Both drugs affect the QT interval of your heart and can lead to arrhythmias and sudden death, especially if you are taking other meds or have a heart condition.} \\
    \textbf{Label:} \lbl{YES} \\
    \textbf{Explanation:} There is a claim in the text.
    \item \textit{Saw this on Facebook today and it’s a must read for all those idiots clearing the shelves \#coronavirus \#toiletpapercrisis \#auspol} \\
    \textbf{Label:} \lbl{NO} \\
    \textbf{Explanation:} There is no claim in the text.
\end{enumerate}

\subsection{False Information}
\uline{\textbf{Question 2:} To what extent does the tweet appear to contain false information?} 

The stated claim may contain false information. This question labels the tweets with the categories mentioned below. \textit{False Information} appears on social media platforms, blogs, and news-articles to deliberately misinform or deceive the readers~\citep{kumar2018false}.

\noindent\textbf{Labels:}
The labels for this question are defined with a five point Likert scale~\citep{albaum1997likert}. A higher value means that it is more likely to be false: 

\begin{enumerate}
\itemsep-0.2em 
\item \lbl{NO, definitely contains no false information}
\item \lbl{NO, probably contains no false information}
\item \lbl{not sure}
\item \lbl{YES, probably contains false information}
\item \lbl{YES, definitely contains false information}
\end{enumerate}


To answer this question, it is recommended to open the link of the tweet and to look for additional information for the veracity of the claim identified in question 1. For example, if the tweet contains a link to an article from a reputable information source (e.g., Reuters, Associated Press, France Press, Aljazeera English, BBC), then the answer could be ``\dots~contains no false info''.

\noindent \textbf{Examples:}
\begin{enumerate}
\item \textit{``Dominican Republic found the cure for Covid-19 https://t.co/1CfA162Lq3''}
\\
\textbf{Label:}\enspace\lbl{5.}\,\lbl{YES, definitely contains false information}
\\
\textbf{Explanation:} This is not correct information at the time of this tweet is posted. 

\item \textit{This is Dr. Usama Riaz. He spent past weeks screening and treating patients with Corona Virus in Pakistan. He knew there was no PPE. He persisted anyways. Today he lost his own battle with coronavirus but he gave life and hope to so many more. KNOW HIS NAME \img{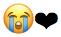} https://t.co/flSwhLCPmx}
\\
\textbf{Label:}\enspace\lbl{2.}\,\lbl{NO, probably contains no false info}
\\
\textbf{Explanation:} The content of the tweet states correct information. 



\end{enumerate}

\subsection{Interest to General Public}
\uline{\textbf{Question 3:} Will the tweet's claim have an effect on or be of interest to the general public?}

Most often people do not make interesting claims, which can be verified by our general knowledge. For example, though \textit{``The sky is blue''} is a claim, it is not interesting to the general public. In general, topics such as healthcare, political news, and current events are of higher interest to the general public. Using the five point Likert scale the labels are defined below.

\noindent\textbf{Labels:}

\begin{enumerate}
\itemsep-0.2em 
\item \lbl{NO, definitely not of interest}
\item \lbl{NO, probably not of interest}
\item \lbl{not sure}
\item \lbl{YES, probably of interest}
\item \lbl{YES, definitely of interest}
\end{enumerate}

\noindent \textbf{Examples:} 
\begin{enumerate}
    \item \textit{Germany is conducting 160k Covid-19 tests a week. It has a total 35k ventilators, 10k ordered to be made by the govt. It has converted a new 1k bed hospital in Berlin.   It’s death rate is tiny bcos it’s mass testing allows quarantine and bcos it has fewer non reported cases.} \\
    \textbf{Label:}\enspace\lbl{4.}\,\lbl{YES: probably of interest} \\
    \textbf{Explanation:} This information is relevant and of high interest for the general population as it reports how a country deals with COVID-19.
    
    
    \item \textit{Fake news peddler Dhruv Rathee had said:  ``Corona virus won't spread outside China, we need not worry''  Has this guy ever spoke something sensible? https://t.co/siBAwIR8Pn} 
    \\
    \textbf{Label:}\enspace\lbl{2.}\,\lbl{NO, probably not of interest}
    \\
    \textbf{Explanation:} The information is not interesting for the general public as it is an opinion and providing statement of someone else.
\end{enumerate}

\subsection{Harmfulness}
\uline{\textbf{Question 4:}  To what extent does the tweet appear to be harmful to society, person(s), company(s) or product(s)?}

The purpose of this question is to determine if the content of the tweet aims to and can negatively affect society as a whole, specific person(s), company(s), product(s), or spread rumors about them. The content intends to harm or \textit{weaponize the information}\footnote{The use of information as a weapon to spread misinformation and mislead people.}~\citep{broniatowski2018weaponized}.
A rumor involves a form of a statement whose veracity is not quickly verifiable or ever confirmed\footnote{\url{https://en.wikipedia.org/wiki/Rumor}}.

\textbf{Labels:} To categorize the tweets we defined the following labels based on the Likert scale. A higher value means a higher degree of harm. 

\begin{enumerate}
\itemsep-0.2em 
\item \lbl{NO, definitely not harmful}
\item \lbl{NO, probably not harmful}
\item \lbl{not sure}
\item \lbl{YES, probably harmful}
\item \lbl{YES, definitely harmful}
\end{enumerate}

\noindent \textbf{Examples: }

\begin{enumerate}
\item \textit{How convenient but not the least bit surprising from Democrats! As usual they put politics over American citizens. @SpeakerPelosi withheld \#coronavirus bill so DCCC could run ads AGAINST GOP candidates! \#tcot} 
\\
\textbf{Label:}\enspace\lbl{5.}\,\lbl{YES, definitely harmful}
\\
\textbf{Explanation:} This tweet is weaponized to target Nancy Pelosi and the Democrats in general.

\item \textit{As we saw over the wkend, disinfo is being spread online about a supposed national lockdown and grounding flights. Be skeptical of rumors. Make sure you’re getting info from legitimate sources. The @WhiteHouse is holding daily briefings and @cdcgov is providing the latest.} 
\\
\textbf{Label:}\enspace\lbl{1.}\,\lbl{NO, definitely not harmful}
\\
\textbf{Explanation:} This tweet is informative and gives advice. It does not attack anyone and is not harmful. 

\end{enumerate}

\subsection{Need of Verification}
\uline{\textbf{Question 5:} Do you think that a professional fact-checker should verify the claim in the tweet?}

It is important to verify a factual claim by a professional fact-checker, as the claim may cause harm to society, specific person(s), company(s), product(s), or some government entities. However, not all factual claims are important or worthwhile to be fact-checked by a professional fact-checker, because it is a time-consuming procedure. Therefore, the purpose is to categorize the tweet using the labels defined below. While doing so, the annotator can rely on the answers to the previous questions. For this question, we defined the following labels to categorize the tweets. 


\noindent \textbf{Labels:}
\begin{enumerate}
\itemsep-0.2em 
\item \lbl{NO, no need to check}: the tweet does not need to be fact-checked, e.g., because it is not interesting, a joke, or does not contain any claim.
\item \lbl{NO, too trivial to check}: the tweet is worth fact-checking, however, this does not require a professional fact-checker, i.e., a non-expert might be able to fact-check the claim. For example, one can verify the information using reliable sources such as the official website of the WHO, etc. An example of a claim is as follows: \textit{``The GDP of the USA grew by 50\% last year.''}
\item \lbl{YES, not urgent}: the tweet should be fact-checked by a professional fact-checker, however, it is not urgent or critical;
\item \lbl{YES, very urgent}: the tweet can cause immediate harm to a large number of people, therefore, it should be verified as soon as possible by a professional fact-checker; 
\item \lbl{Not sure}: the content of the tweet does not have enough information to make a judgment. 
\end{enumerate}

\noindent\textbf{Examples:} 
\begin{enumerate}
\item \textit{Things the GOP has done during the Covid-19 outbreak:   - Illegally traded stocks  - Called it a hoax - Blamed it on China  - Tried to bailout big business without conditions  What they haven’t done:   - Help workers  - Help small businesses - Produced enough tests or ventilators}
\\
\textbf{Label:}\enspace\lbl{2.}\,\lbl{YES, very urgent}
\\
\textbf{Explanation:} Clearly, the content of the tweet blames authority, hence, it is important to verify this claim immediately by a professional fact-checker. In addition, the attention of government entities might be required in order to take necessary actions.
\item \textit{ALERT \img{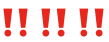} The corona virus can be spread through internationally printed albums. If you have any albums at home, put on some gloves, put all the albums in a box and put it outside the front door tonight. I'm collecting all the boxes tonight for safety. Think of your health.}
\\
\textbf{Label:}\enspace\lbl{5.}\,\lbl{NO, no need to check}
\\
\textbf{Explanation:} This is a joke and does not need to be checked by a professional fact checker. 
\end{enumerate}

\subsection{Harmful to Society}
\uline{\textbf{Question 6:} Is the tweet harmful for society and why?}

The purpose of this question is to categorize if the content of the tweet is intended to harm or is weaponized to mislead the society. To identify that we defined the following labels for the categorization. 

\noindent\textbf{Labels:}
\begin{enumerate}[label=\Alph*.]
\itemsep-0.2em 
\item \lbl{NO, not harmful:} the content of the tweet would not harm the society (e.g., \textit{``I like corona beer''}).
\item \lbl{NO, joke or sarcasm:} the tweet contains a joke (e.g., \textit{``If Corona enters Spain, it’ll enter from the side of Barcelona defense''}) or sarcasm (e.g., \textit{```The corona virus is a real thing.' -- Wow, I had no idea!''}).
\item \lbl{not sure:} if the content of the tweet is not understandable enough to judge.
\item \lbl{YES, panic:} the tweet spreads panic. The content of the tweet can cause sudden fear and anxiety for a large part of the society (e.g., \textit{``there are 50,000 cases ov COVID-19 in Qatar''}).
\item \lbl{YES, xenophobic, racist, prejudices, or hate-speech:} the tweet reports xenophobia, racism or prejudiced expression(s). 
According to the dictionary\footnote{\url{https://www.dictionary.com/}} \textit{Xenophobic} refers to fear or hatred of foreigners, people from different cultures, or strangers. \textit{Racism} is the belief that groups of humans possess different behavioral traits corresponding to physical appearance and can be divided based on the superiority of one race over another.\footnote{\url{https://en.wikipedia.org/wiki/Racism}} It may also refer to prejudice, discrimination, or antagonism directed against other people because they are of a different race or ethnicity. \textit{Prejudice} is an unjustified or incorrect attitude (i.e., typically negative) towards an individual based solely on the individual's membership of a social group.\footnote{\url{https://www.simplypsychology.org/prejudice.html}}
An example of a xenophobic statement is \textit{``do not buy cucumbers from Iran''}.
\item \lbl{YES, bad cure:} the tweet reports a questionable cure, medicine, vaccine or prevention procedures (e.g., \textit{``\dots drinking bleach can help cure coronavirus''}). 

\item \lbl{YES, rumor, or conspiracy:} the tweet reports or spreads a rumor. It is defined as a ``specific (or topical) proposition for belief passed along from person to person usually by word of mouth without secure standards of evidence being present''~\citep{allport1947psychology}.  
For example, \textit{``BREAKING: Trump could still own stock in a company that, according to the CDC, will play a major role in providing coronavirus test kits to the federal government, which means that Trump could profit from coronavirus testing. \#COVID-19 \#coronavirus https://t.co/Kwl3ylMZRk''}
\item \lbl{YES, other:} if the content of the tweet does not belong to any of the above categories, then this category can be chosen to label the tweet.

\end{enumerate}


\subsection{Requires attention}
\uline{\textbf{Question 7:} Do you think that this tweet should get the attention of any government entity?}

Most often people tweet by blaming authorities, providing advice, and/or call for action. Sometimes that information might be useful for some government entities to make a plan, respond or react on it. The purpose of this question is to categorize such information. It is important to note that not all information requires attention from a government entity. Therefore, even if the tweet's content belongs to any of the positive categories, it is important to understand whether that requires government attention. For the annotation, it is mandatory to first decide on whether attention is necessary from 
government entities (i.e., \lbl{YES/NO}). If the answer is \lbl{YES}, it is obligatory to select a category from the \lbl{YES} sub-categories mentioned below.


\noindent\textbf{Labels:}
\begin{enumerate}[label=\Alph*.]
\itemsep-0.2em 
\item \lbl{NO, not interesting:} if the content of the tweet is not important or interesting for any government entity to pay attention to. 
\item \lbl{not sure:} if the content of the tweet is not understandable enough to judge. 
\item \lbl{YES, categorized as in question 6:} if some government entities need to pay attention to this tweet as it is harmful for society, i.e., it is labeled as any of the \textit{YES} sub-categories in question 6. 
\item \lbl{YES, other:} if the tweet cannot be labeled as any of the above categories, then this label should be selected. 
\item \lbl{YES, blame authorities:} the tweet contains information that blames some government entities or top politician(s), e.g., \textit{``Dear @VP Pence: Is the below true? Do you have a plan? Also, when are local jurisdictions going to get the \#Coronavirus test kits you promised?''}.

\item \lbl{YES, contains advice:} the tweet contains advice about social, political, national, or international issues that requires attention from some government entities (e.g., \textit{The elderly \& people with pre-existing health conditions are more susceptible to \#COVID19. To stay safe, they should: \checkmark Keep distance from people who are sick \checkmark Frequently wash hands with soap \& water \checkmark  Protect their mental health}).

\item \lbl{YES, calls for action:} the tweet contains information that states that some government entities should take action for a particular issue (e.g., \textit{I think the Government should close all the Barber Shops and Salons , let people buy shaving machines and other beauty gardgets keep in their houses. Salons and Barbershops might prove to be another Virus spreading channels @citizentvkenya @SenMutula @CSMutahi\_Kagwe}).

\item \lbl{YES, discusses action taken:} if the tweet discusses actions taken by governments, companies, individuals for any particular issue, for example, closure of bars, conferences, churches due to the corona virus (e.g., \textit{Due to the current circumstances with the Corona virus, The 4th Mediterranean Heat Treatment and Surface Engineering Conference in Istanbul postponed to 26-28 Mayıs 2021.}). 

\item \lbl{YES, discusses cure:} if attention is needed from some government entities as the tweet discusses a possible cure, vaccine, or treatment for a disease (e.g., \textit{Pls share this valuable information. Garlic boiled water can be cure corona virus}). 

\item \lbl{YES, asks question:} if the content of the tweet contains a question over a particular issue and it requires attention from government entities (e.g., \textit{Special thanks to all doctors and nurses, new found respect for you’ll. Is the virus going to totally disappear in the summer? I live in USA and praying that when the temperature warms up the virus will go away...is my thinking accurate?})

\end{enumerate}

\subsection{Multimedia in Tweets}
In this subsection, we study the correlation between tweet labels and multimedia (video, image, or none). Generally, people trust videos more than images or plain texts which suggests that tweets with video potentially have a higher impact.

In the Arabic dataset, we didn't find any clear preference for Q1-Q4, i.e., a tweet with video, image, or only text can contain a factual claim with almost similar ratios. For Q5 (i.e., need fact-checking by a professional fact-checker), when a tweet has a video, in 44\% of the cases, annotators selected ``Yes'' compared to 17\% and 25\% for image and text only respectively. Fact-checking of videos is not always trivial. For Q6 (i.e., harmful to society) and Q7 (i.e., should attract government attention), when a tweet has a video, it has the potential to be harmful and get government attention is higher than tweets with only images or text. Annotators selected ``Yes'' in almost 33\% of the tweets having videos for Q6 and Q7 and this ratio decreased to almost half for tweets having images or text only. This shows the importance of having videos in tweets as it gives more trust.

\subsection{Geographical distribution}
\begin{figure} 
\centering
    \begin{subfigure}[b]{0.45\textwidth}
        \includegraphics[width=\textwidth]{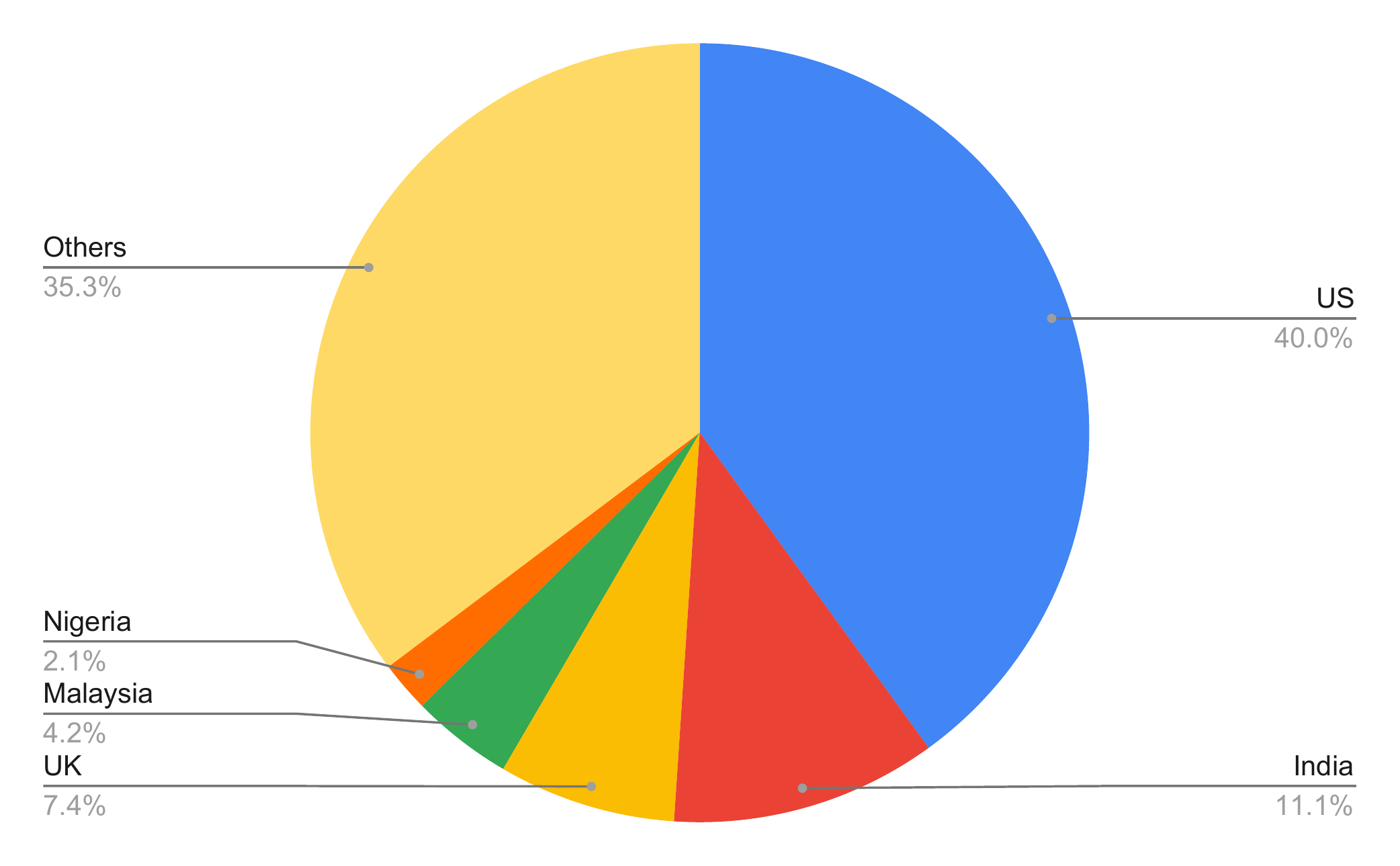}
        \caption{English dataset}
        \label{fig:countries_en}
    \end{subfigure}
    \begin{subfigure}[b]{0.45\textwidth}    
        \includegraphics[width=\textwidth]{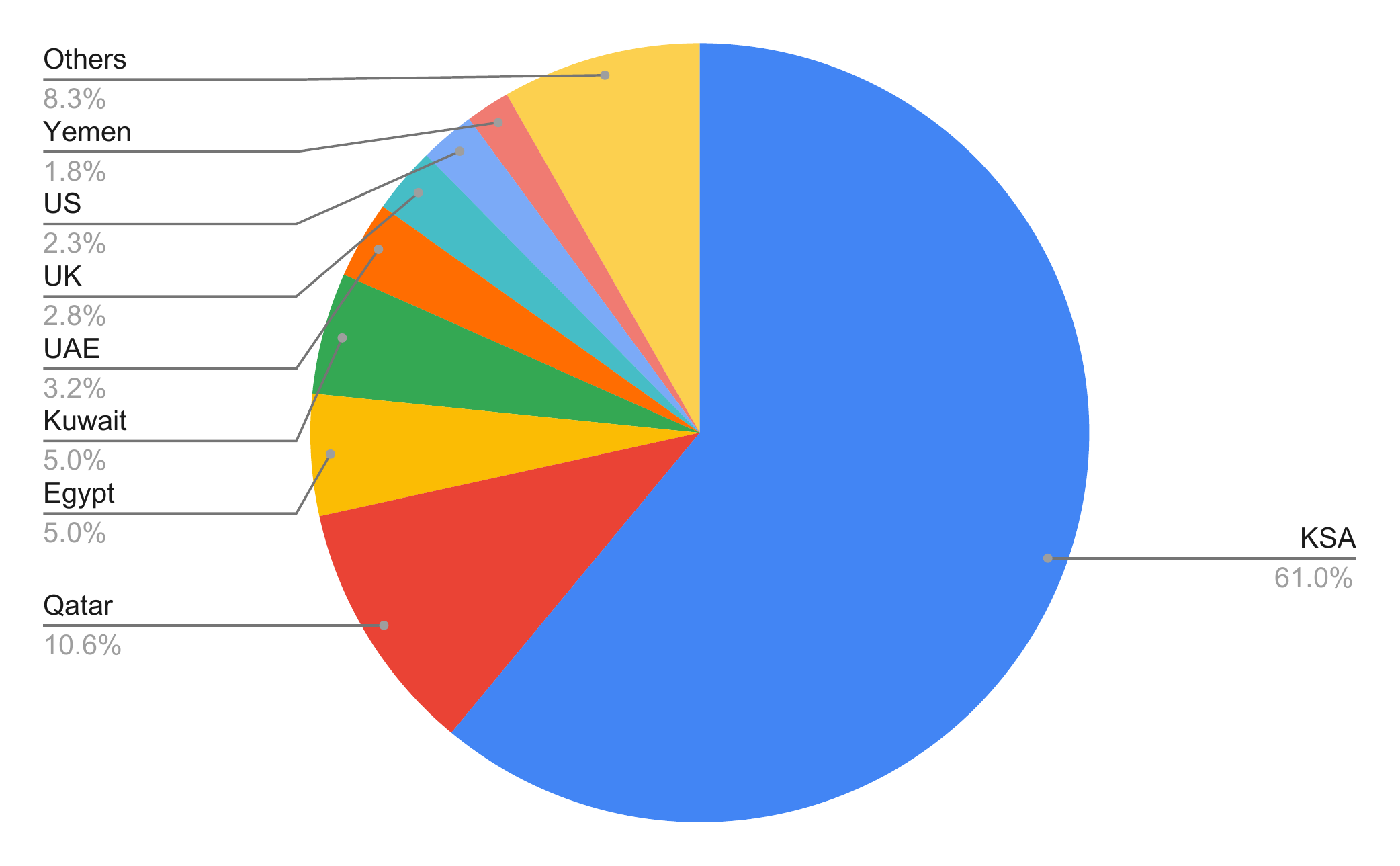}
        \caption{Arabic dataset}
        \label{fig:countries_en_ara}    
    \end{subfigure} 
\caption{Country distribution for English and Arabic tweets}
\label{fig:tweets_country_dist}
\end{figure}

Figure \ref{fig:tweets_country_dist} shows the geographical distribution of annotated tweets for English and Arabic. We consider the country of the tweet author or the original author in case of retweeting. It is observed that most English tweets came from the US, India, and the UK ($\sim$60\%), while most Arabic tweets came from KSA and Qatar ($\sim$70\%). For both languages, there are tweets from a large number of countries, which indicates a good diversity of interests, topics, styles, etc. that strengthens our study.

\begin{figure*}[t]
\centering
\includegraphics[width=\textwidth]{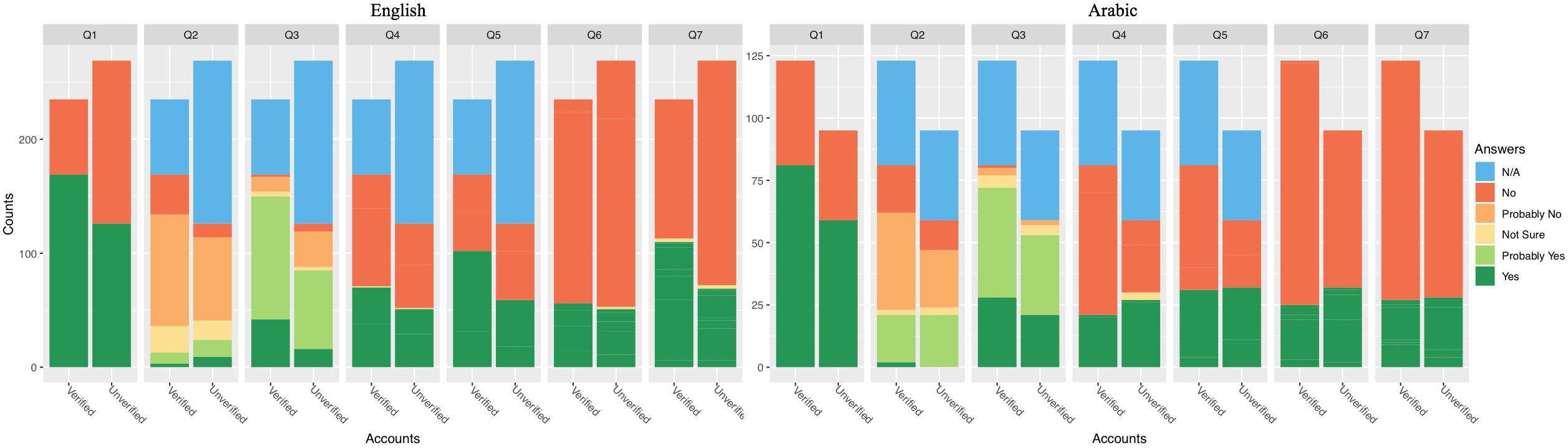}
\caption{Distribution of datasets for all the questions associated with user accounts. NA refers to tweets that have not been labeled for those questions, they are identical to the tweets categorized with the label NO in Q1.}
\label{fig:user_account_label_dist}
\end{figure*}

\subsection{Verified and Unverified Accounts}
We study the correlation between tweet labels and whether or not the original author of a tweet has a verified account. Verified accounts include government entities, public figures, celebrities, etc., which have a large number of followers, so their tweets typically have a high impact on society.

Figure \ref{fig:user_account_label_dist} shows that verified accounts tend to post more tweets that contain factual claims than unverified accounts (Q1), and their tweets are more likely to not contain false information (Q2), be of higher interest to the general public (Q3), be less harmful to society (Q6, Arabic), and attract greater attention from a government entity than tweets from unverified accounts (Q7, English). These are general observations from the currently small number of annotated tweets, and there are some differences between the English and Arabic annotations. The quantitative study can be held at a later stage using a larger dataset.

This correlation could be one of the features that a classifier can use to predict labels for unseen tweets, can also help in speeding up the annotation process by providing initial default values before manual revision. In addition, in some cases, verified accounts can be used to check annotation quality, for example, tweets from @WHO should not be labeled as weaponized or harmful to society.

\subsection{Correlation Between Questions}
\label{sec:correlation}

\begin{figure*}
\centering
    \begin{subfigure}[b]{0.5\textwidth}
        \includegraphics[width=\textwidth]{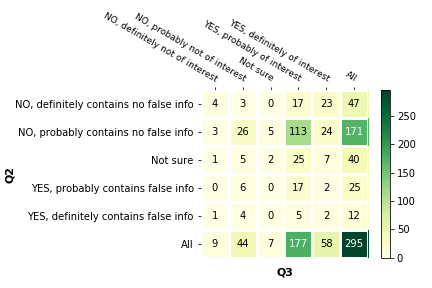}
        \caption{Heatmap for Q2 and Q3.}
        \label{fig:contingency_table_q2_q3}
    \end{subfigure}%
    \begin{subfigure}[b]{0.5\textwidth}    
        \includegraphics[width=\textwidth]{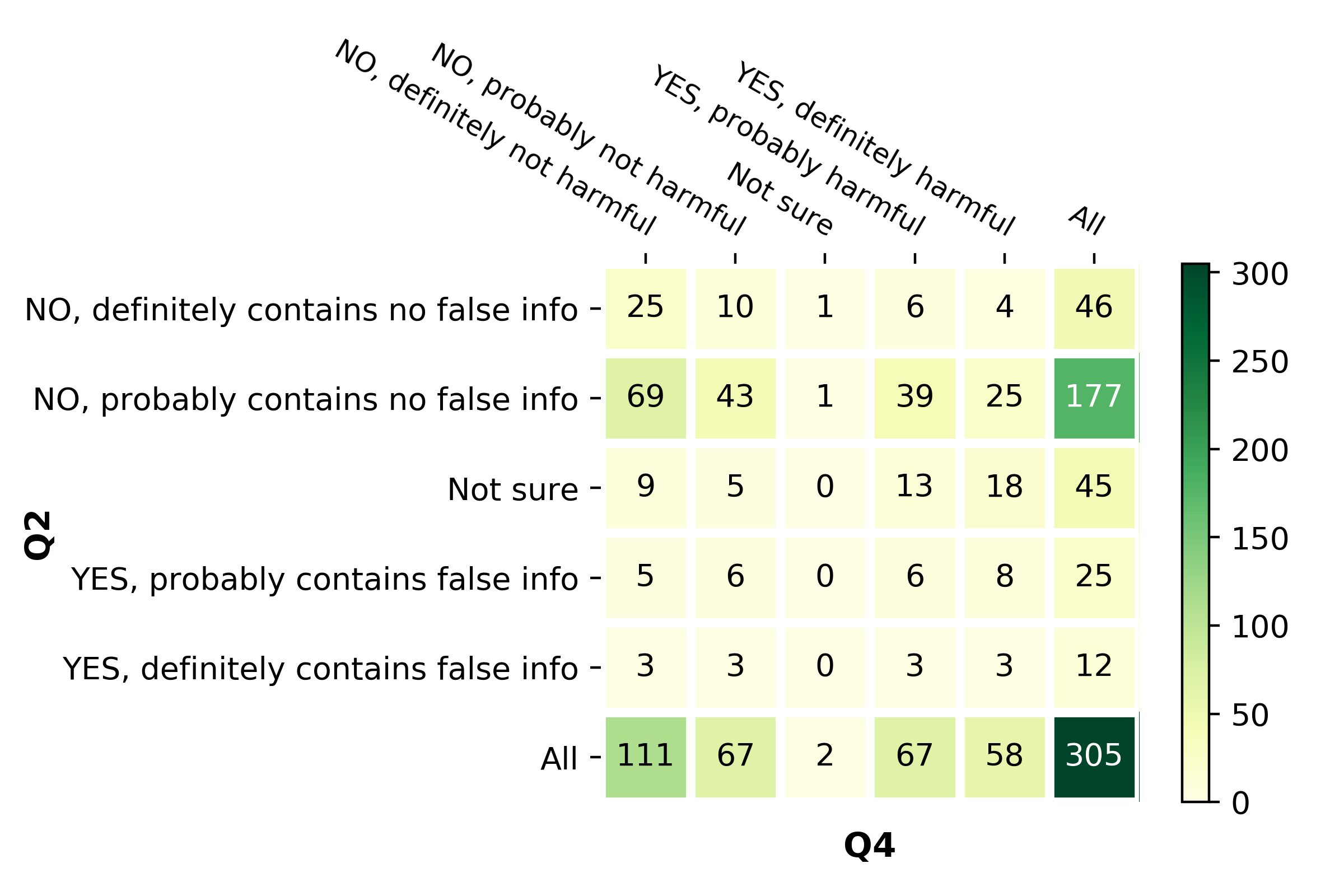}
        \caption{Heatmap for Q2 and Q4.}
        \label{fig:contingency_table_q2_q4}    
    \end{subfigure} 
    \hfill
    \begin{subfigure}[b]{0.5\textwidth}    
        \includegraphics[width=\textwidth]{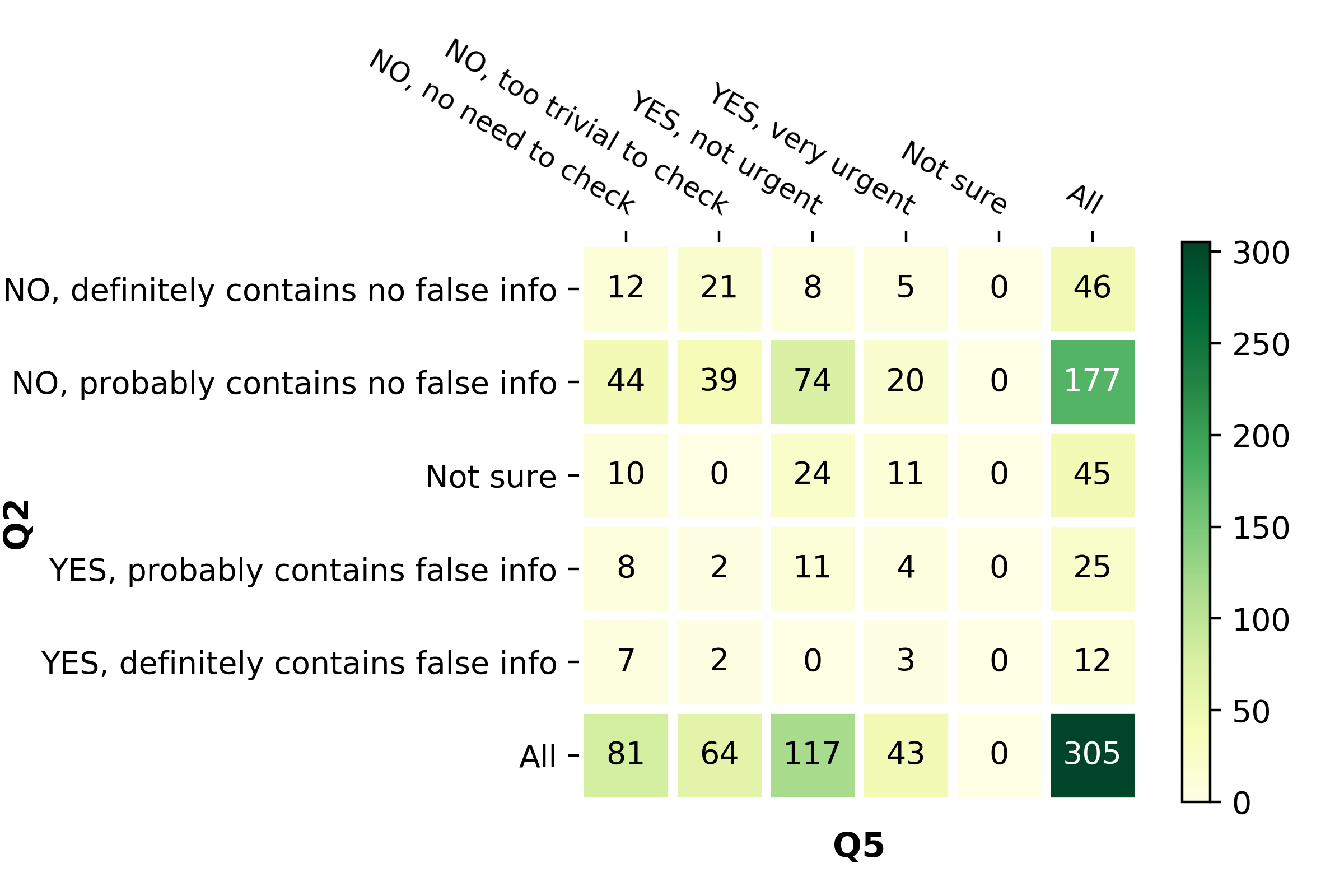}
        \caption{Heatmap for Q2 and Q5.}
        \label{fig:contingency_table_q2_q5}    
    \end{subfigure}%
    \begin{subfigure}[b]{0.5\textwidth}    
        \includegraphics[width=\textwidth]{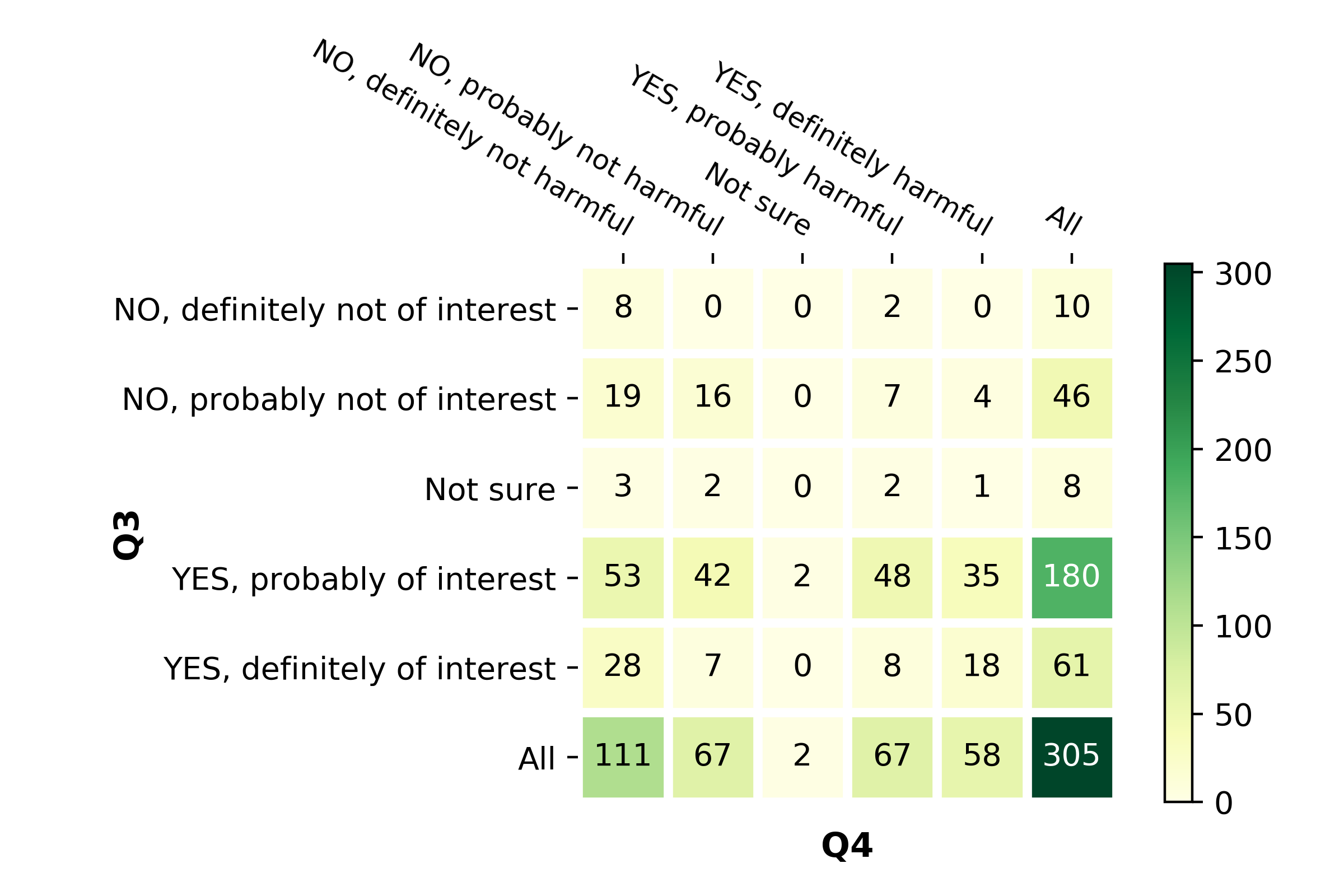}
        \caption{Heatmap for Q3 and Q4.}
        \label{fig:contingency_table_q3_q4}    
    \end{subfigure}
    \hfill
    \begin{subfigure}[b]{0.5\textwidth}    
        \includegraphics[width=\textwidth]{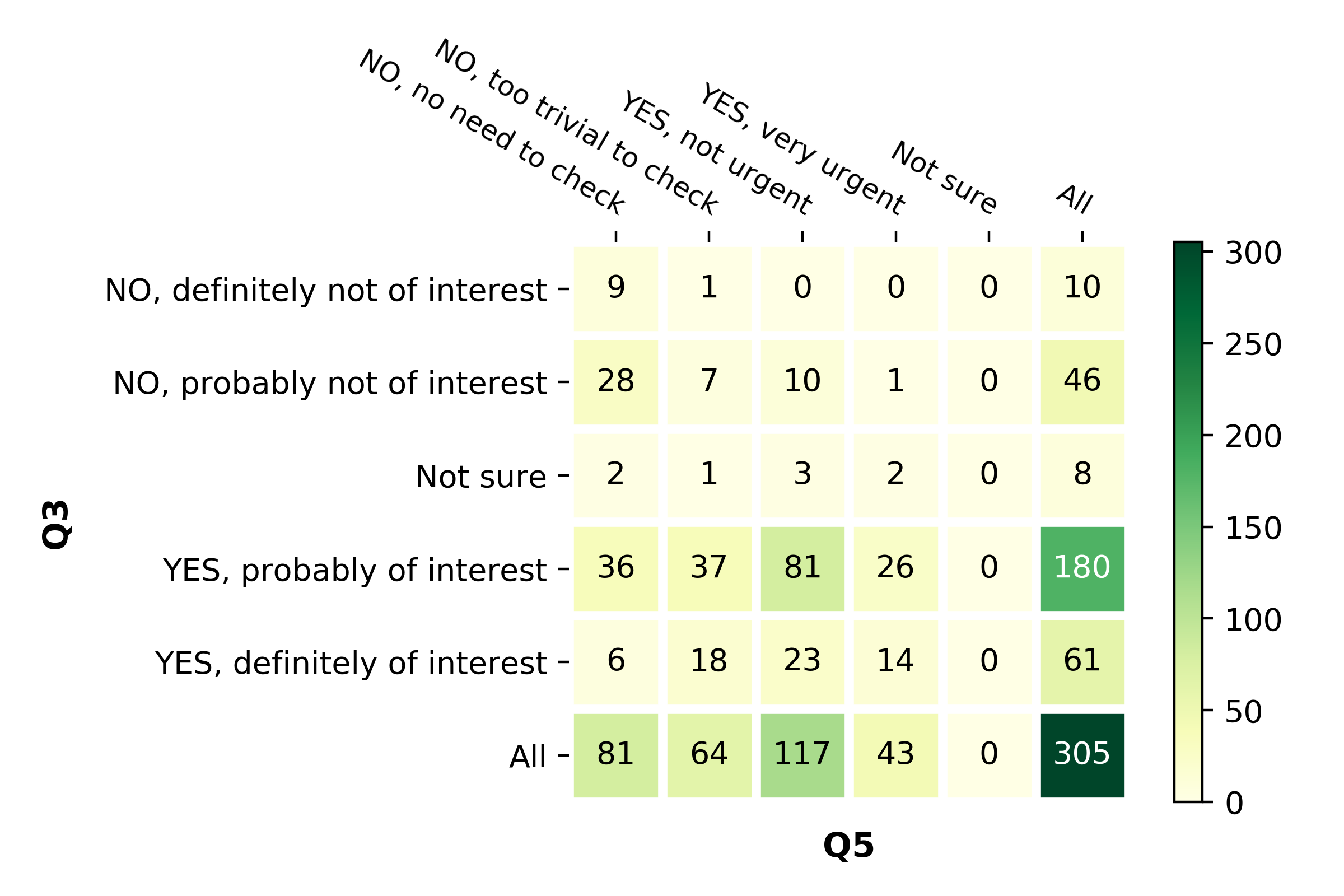}
        \caption{Heatmap for Q3 and Q5.}
        \label{fig:contingency_table_q3_q5}    
    \end{subfigure}%
    \begin{subfigure}[b]{0.5\textwidth}    
        \includegraphics[width=\textwidth]{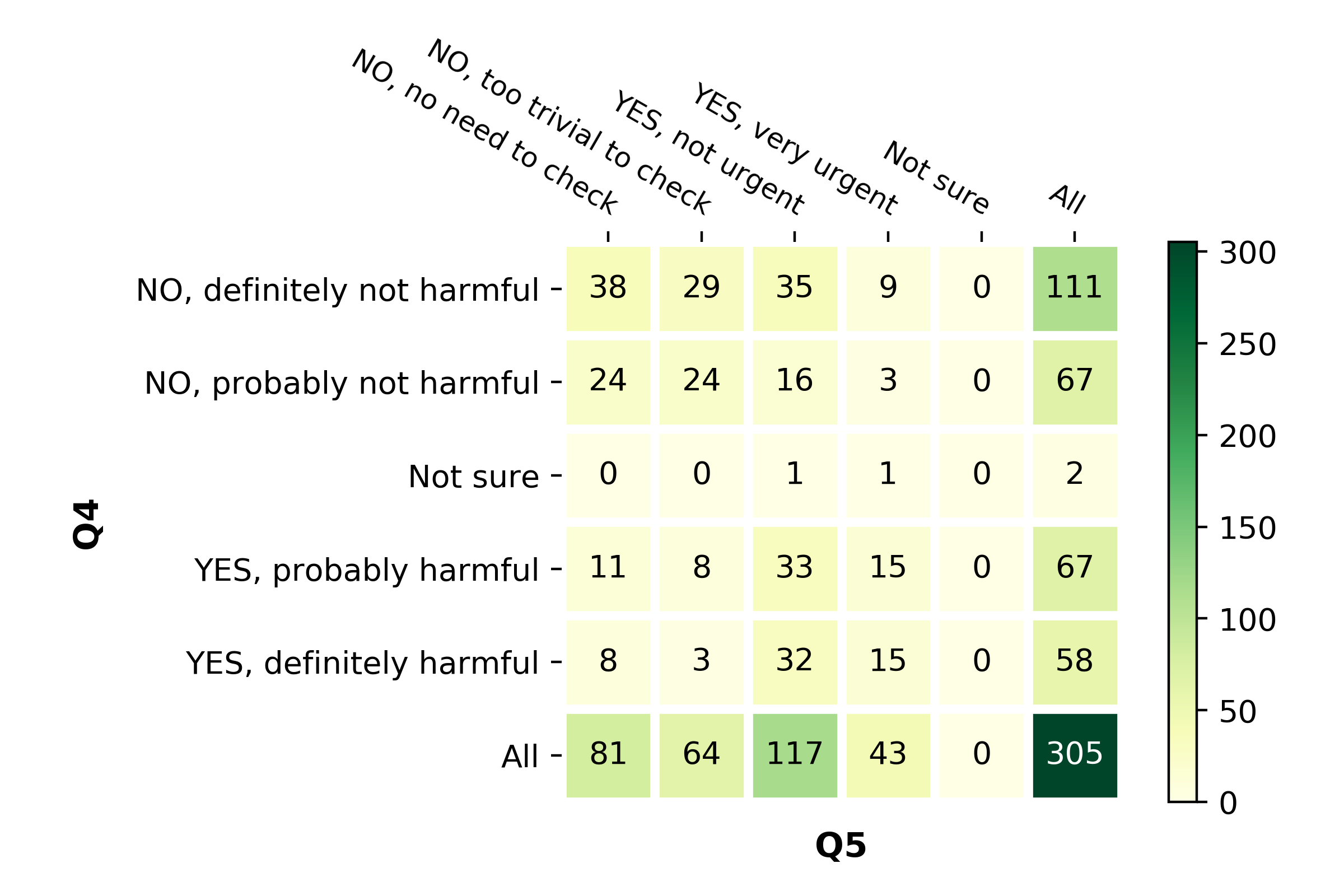}
        \caption{Heatmap for Q4 and Q5.}
        \label{fig:contingency_table_q4_q5}    
    \end{subfigure}     
    \hfill
    \begin{subfigure}[b]{0.5\textwidth}    
        \includegraphics[width=\textwidth]{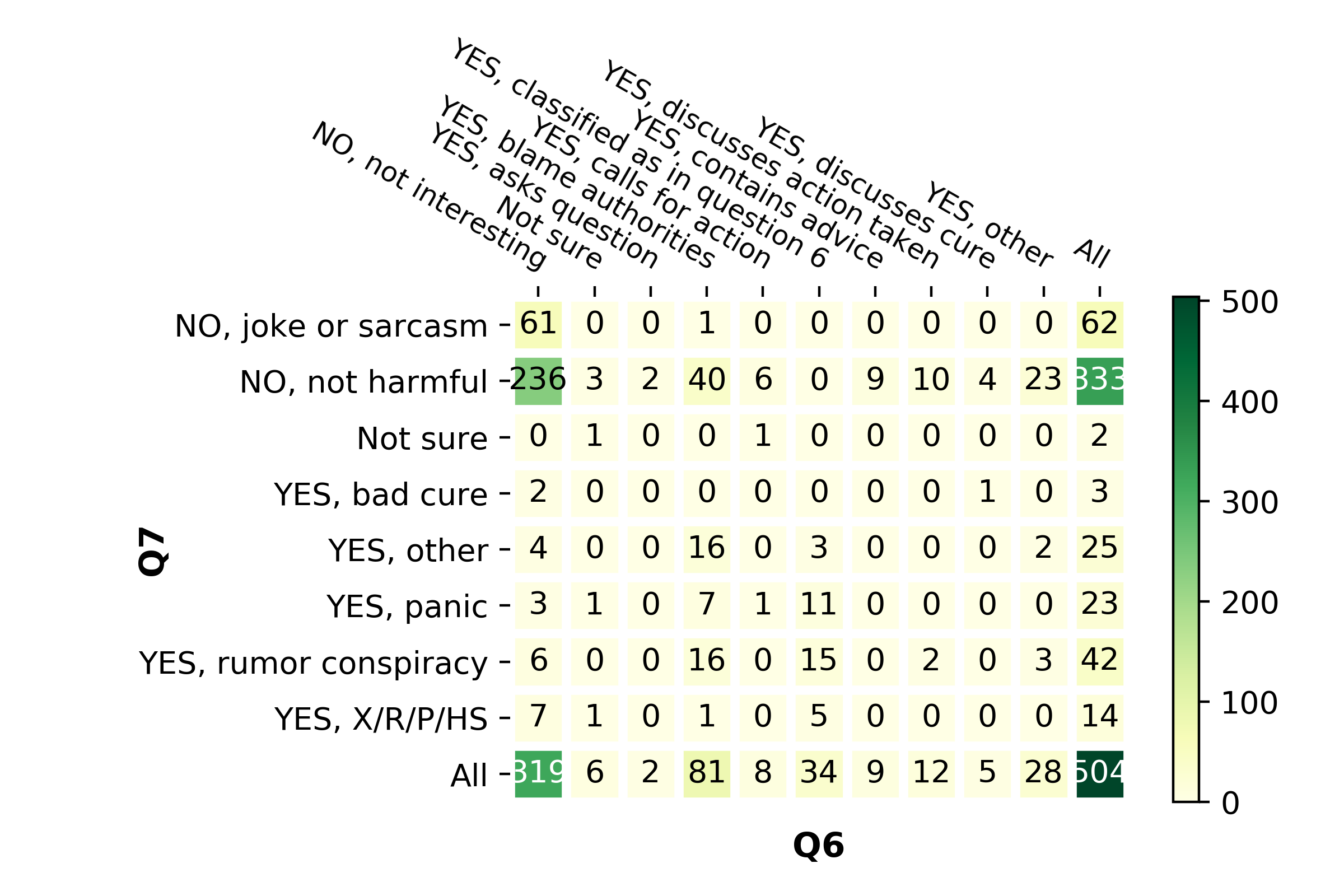}
        \caption{Heatmap for Q6 and Q7. YES, X/R/P/HS -- YES, xenophobic, racist, prejudices or hate speech}
        \label{fig:contingency_table_q6_q7}    
    \end{subfigure}%
    \begin{subfigure}[b]{0.5\textwidth}    
            
        \includegraphics[width=\textwidth]{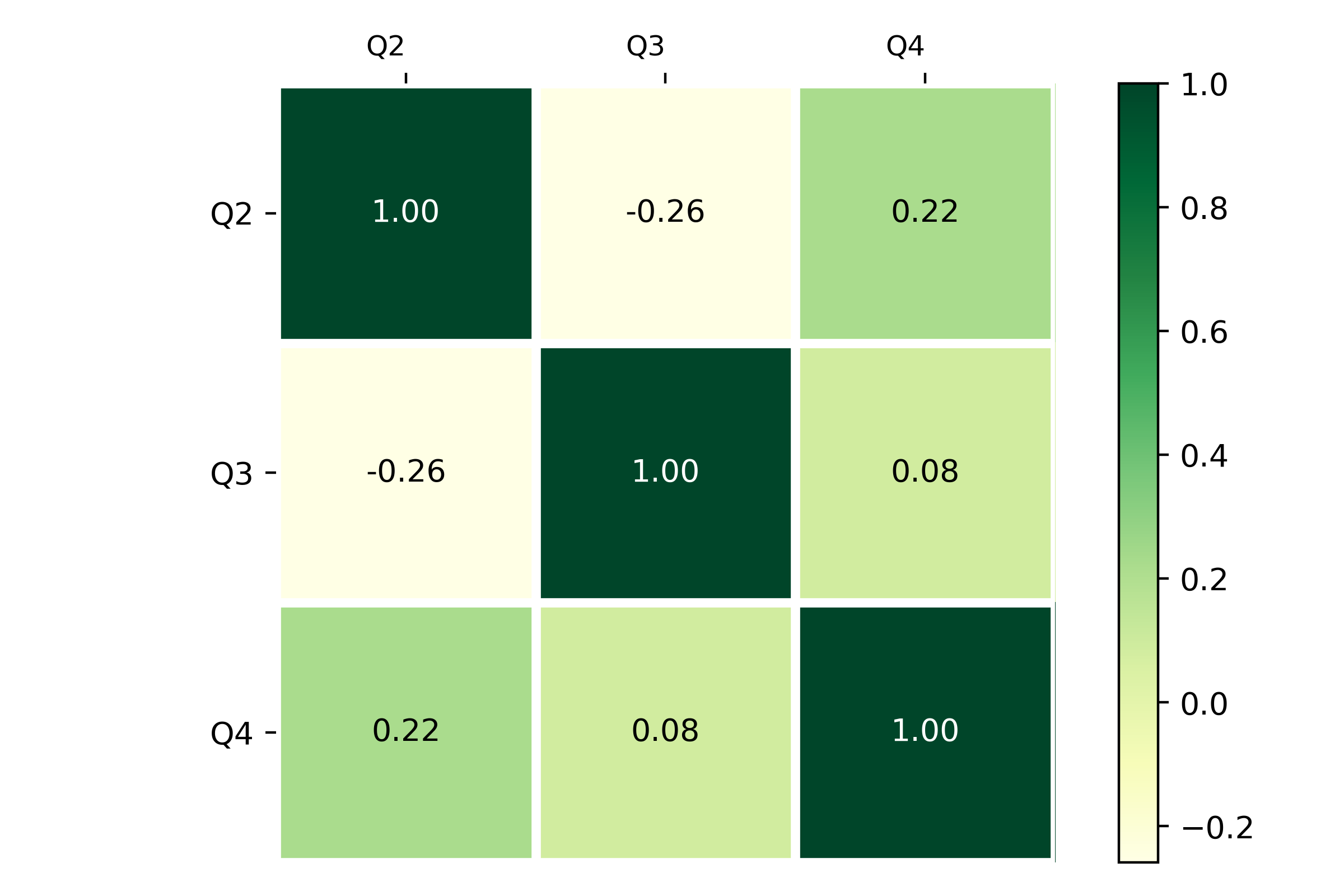}
        \caption{Correlation between Q2 to Q4.}
        \label{fig:english_correlation_q2_q4}

    \end{subfigure}    
    \caption{Contingency and correlation heatmaps of \textbf{English tweets} for different question pairs}
    \label{fig:contingency_all}
\end{figure*}

\begin{figure*} 
\centering
    \begin{subfigure}[b]{0.5\textwidth}
        \includegraphics[width=\textwidth]{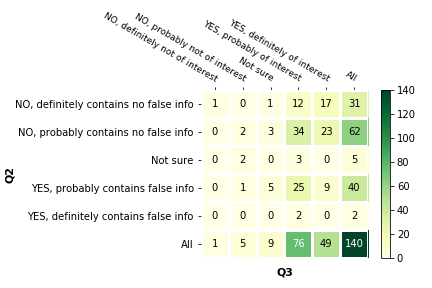}
        \caption{Heatmap for Q2 and Q3.}
        \label{fig:arabic_contingency_table_q2_q3}
    \end{subfigure}%
    \begin{subfigure}[b]{0.5\textwidth}    
        \includegraphics[width=\textwidth]{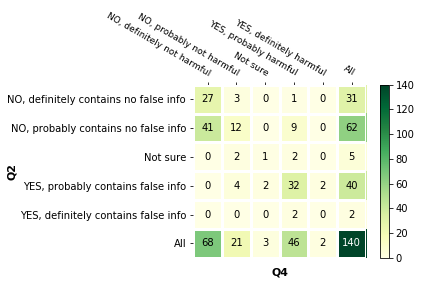}
        \caption{Heatmap for Q2 and Q4.}
        \label{fig:arabic_contingency_table_q2_q4}    
    \end{subfigure} 
    \hfill
    \begin{subfigure}[b]{0.5\textwidth}    
        \includegraphics[width=\textwidth]{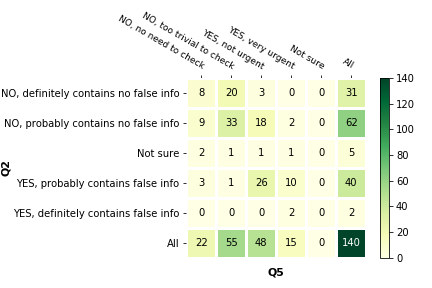}
        \caption{Heatmap for Q2 and Q5.}
        \label{fig:arabic_contingency_table_q2_q5}    
    \end{subfigure}%
    \begin{subfigure}[b]{0.5\textwidth}    
        \includegraphics[width=\textwidth]{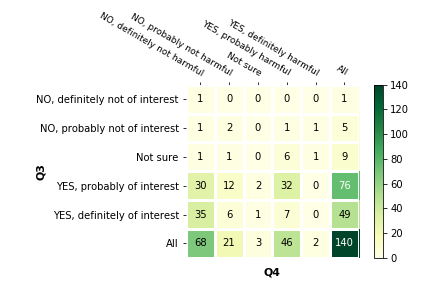}
        \caption{Heatmap for Q3 and Q4.}
        \label{fig:arabic_contingency_table_q3_q4}    
    \end{subfigure}
    \hfill
    \begin{subfigure}[b]{0.5\textwidth}    
        \includegraphics[width=\textwidth]{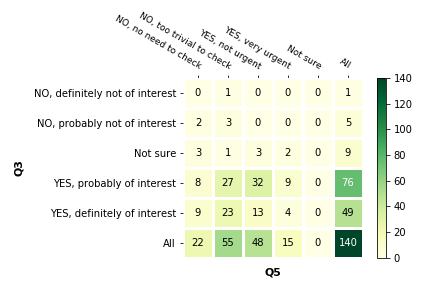}
        \caption{Heatmap for Q3 and Q5.}
        \label{fig:arabic_contingency_table_q3_q5}    
    \end{subfigure}%
    \begin{subfigure}[b]{0.5\textwidth}    
        \includegraphics[width=\textwidth]{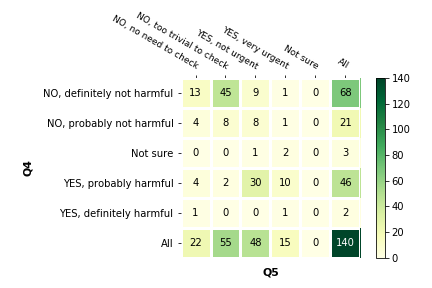}
        \caption{Heatmap for Q4 and Q5.}
        \label{fig:arabic_contingency_table_q4_q5}    
    \end{subfigure}     
    \hfill
    \begin{subfigure}[b]{0.5\textwidth}    
        \includegraphics[width=\textwidth]{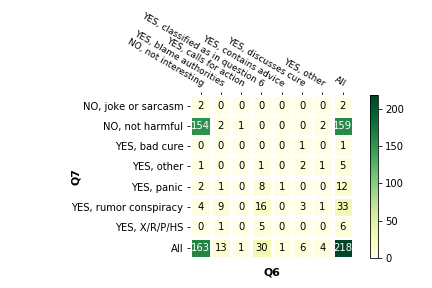}
        \caption{Heatmap for Q6 and Q7. YES, X/R/P/HS -- YES, xenophobic, racist, prejudices or hate speech}
        \label{fig:arabic_contingency_table_q6_q7} 
        
    \end{subfigure}%
    \begin{subfigure}[b]{0.5\textwidth}    
        \includegraphics[width=\textwidth]{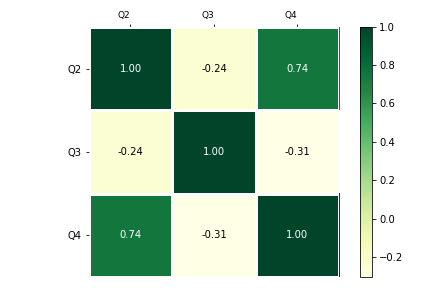}
        \caption{Correlation between Q2 to Q4.}
        \label{fig:arabic_correlation_q2_q4}
    \end{subfigure}    
    \caption{Contingency and correlation heatmaps of \textbf{Arabic tweets} for different question pairs }
    \label{fig:arabic_contingency_all}
\end{figure*}

\subsubsection{English Tweets Dataset}
In Figure \ref{fig:contingency_all}, we report the contingency and correlation tables in a form of a heatmap for different question pairs obtained from the English tweet dataset. For questions Q2-3, it appears that there is a high association
between ``\dots no false info'' and the general public interest as shown in Figure  \ref{fig:contingency_table_q2_q3}. For questions Q2 and Q4 
(Figure \ref{fig:contingency_table_q2_q4}), a high association can be observed between ``\dots~no false info'' and ``\dots~not harmful'' (65\%) compared to ``harmful'' (34\%) for either an individual, products or government entities. By analyzing questions Q2 and Q5 (Figure \ref{fig:contingency_table_q2_q5}), we conclude that ``\dots no false info'' is associated with either ``no need to check'' or ``too trivial to check'', highlighting the fact that a professional fact-checker does not need to spend time on them. From questions Q3 and Q4 (Figure \ref{fig:contingency_table_q3_q4}), it appears that when the content of the tweets is ``not harmful'' the general public interest is higher (61\%) than when it is ``harmful'' (39\%).
From question Q3 and Q5 (Figure \ref{fig:contingency_table_q3_q5}), we see an interesting phenomenon, namely tweets with a high general public interest have a greater association with a professional fact-checker having to verify them (61\%) compared to either ``too trivial to check'' or ``no need to check'' (39\%). The questions Q4 and Q5 (Figure \ref{fig:contingency_table_q4_q5}) show that ``harmful'' tweets require more attention (53\%) from a professional fact-checkers than ``not harmful'' tweets (45\%). Our findings for Q6 and Q7 (Figure \ref{fig:contingency_table_q6_q7}) suggest that the majority of the tweets are not harmful for society, which also requires less attention from government entities. The second most common tweet label for Q7 blames authorities, though they are mostly not harmful for society.    

In Figure \ref{fig:english_correlation_q2_q4}, we report the correlation between questions Q2-4 for the English tweets in order to understand their association. We computed the correlation using the Likert scale values (i.e., 1-5) that we defined for these questions. We observed that overall Q2 and Q3 are negatively correlated, which infers that if the claim contains no false information, it is of high interest to the general public. This can be also observed in Figure \ref{fig:contingency_table_q2_q3}. Questions Q2 and Q4 show a positive correlation, which might be due to their high association with ``\dots no false info'' and ``\dots not harmful''.



\subsubsection{Arabic Tweets Dataset}
In Figure \ref{fig:arabic_contingency_all}, we report heatmaps to illustrate the association across questions using the Arabic tweets. From Q2 and Q3 (Figure \ref{fig:arabic_contingency_table_q2_q3}), we can observe that the association between ``\dots contains no false info'' and general public interest is higher (67\%) than ``\dots contains false info'' (29\%). From questions Q2 and Q4 (Figure \ref{fig:arabic_contingency_table_q2_q4}), we conclude that ``\dots contains no false info'' is associated with ``\dots not harmful'' and ``\dots contains false info'' is associated with ``\dots harmful'', which can also be established from its high correlation of 0.74 in Figure \ref{fig:arabic_correlation_q2_q4}. From the relation between Q2 and Q5 (Figure \ref{fig:arabic_contingency_table_q2_q5}), it can be seen that in the majority of the cases ``\dots contains no false info'' is associated with either ``no need to check'' or ``too trivial to check'', which means that a professional fact-checker does not need to verify them. The analysis between questions Q3 and Q4 suggests that general public interest is higher when the content of the tweets is not harmful (68\%) than harmful (30\%) (Figure \ref{fig:arabic_contingency_table_q3_q4}). From questions Q3 and Q5, we can observe that the general public interest is higher when the claim(s) in the tweets are either ``no need to check'' or ``too trivial to check'' (Figure \ref{fig:arabic_contingency_table_q3_q5}). 
The analysis between question Q4 and Q5 shows that ``not harmful'' tweets are either ``no need to check'' or ``too trivial to check'' by a professional fact-checker (Figure \ref{fig:arabic_contingency_table_q4_q5}). From the questions Q6 and Q7, we notice that in the majority of the cases the tweets are not harmful for society and hence they are not interesting for government entities (Figure \ref{fig:arabic_contingency_table_q6_q7}).


\subsection{Class Label Distribution}

\begin{figure*} 
\centering
        \begin{subfigure}[b]{0.75\textwidth}
        \includegraphics[width=\textwidth]{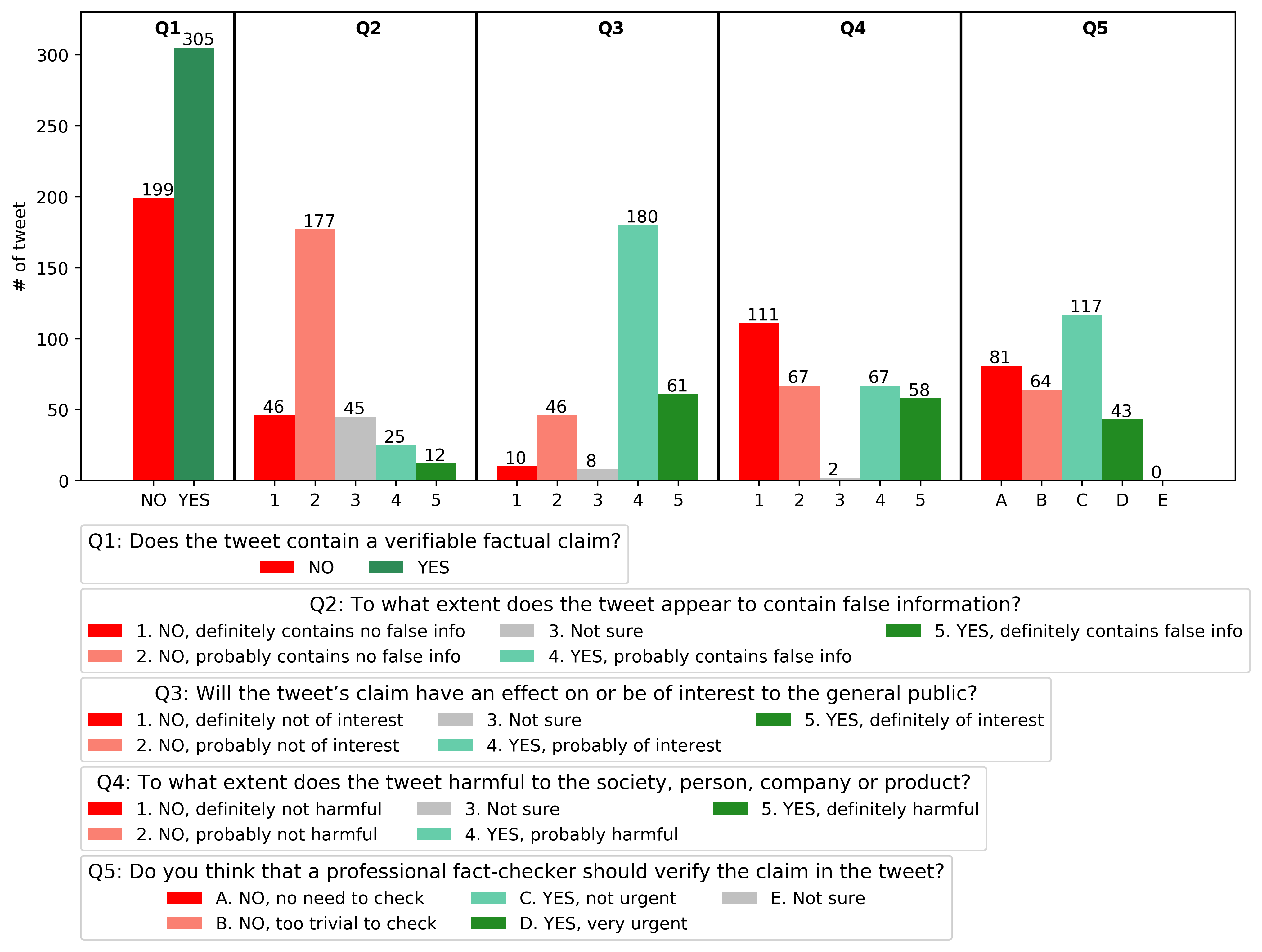}
        \caption{Questions (Q1-5).}
        \label{fig:data_dist_group1}
    \end{subfigure}
    \hfill
    \begin{subfigure}[b]{0.75\textwidth}
        \includegraphics[width=\textwidth]{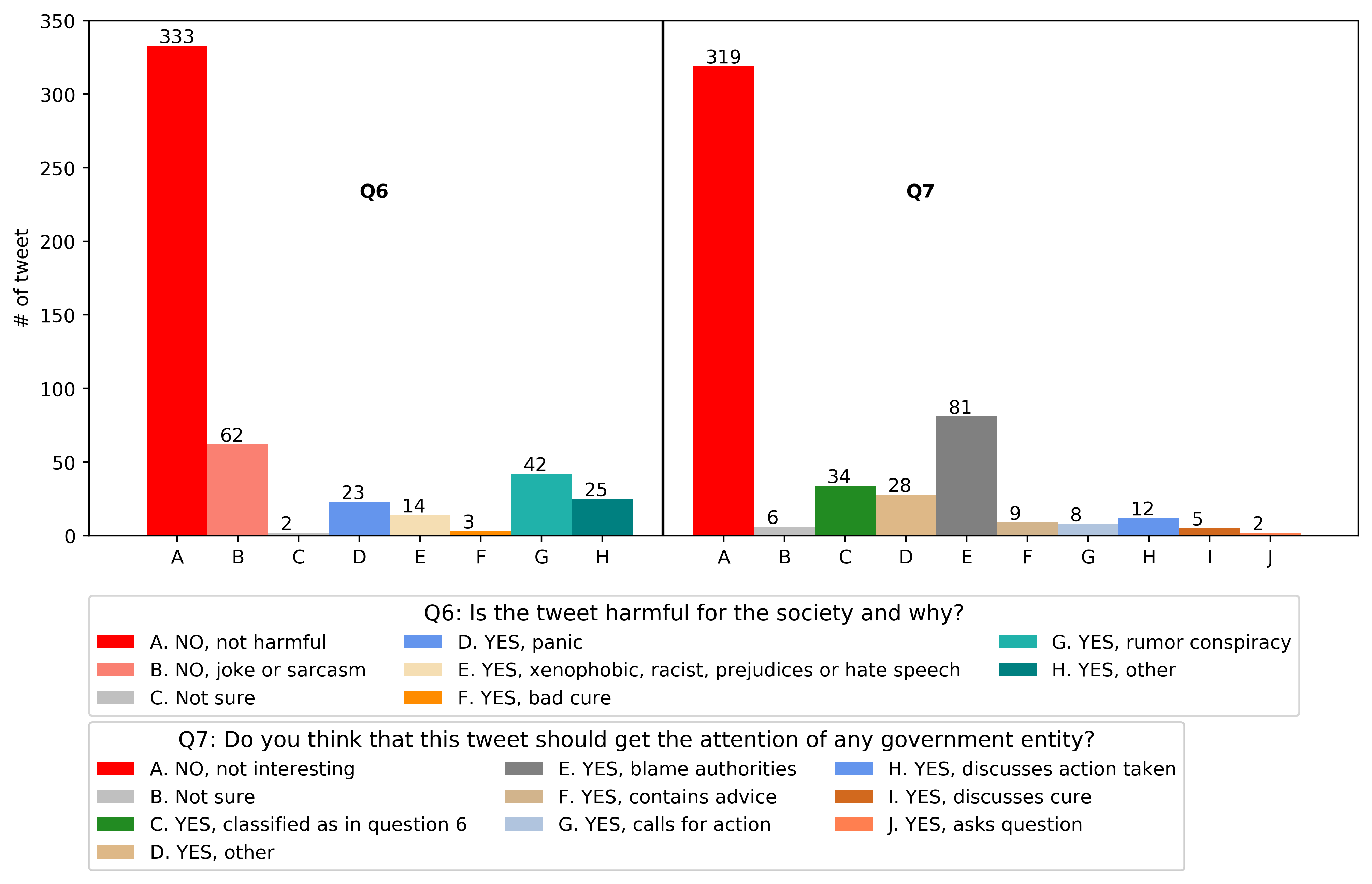}
        \caption{Questions (Q6-7).}
        \label{fig:data_dist_group2}
    \end{subfigure}
    \caption{Distribution of class labels for \textbf{English tweets}}
    \label{fig:data_dist_english_all}
\end{figure*}

\begin{figure*} 
\centering
    \begin{subfigure}[b]{0.75\textwidth}
        \includegraphics[width=\textwidth]{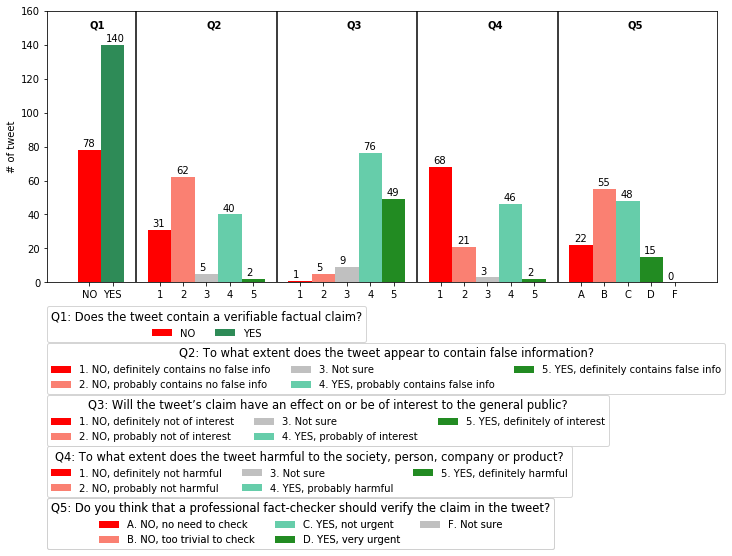}
        \caption{Questions (Q1-5).}
        \label{fig:data_dist_group1_arabic}
    \end{subfigure}
    \hfill
    \begin{subfigure}[b]{0.75\textwidth}    
        \includegraphics[width=\textwidth]{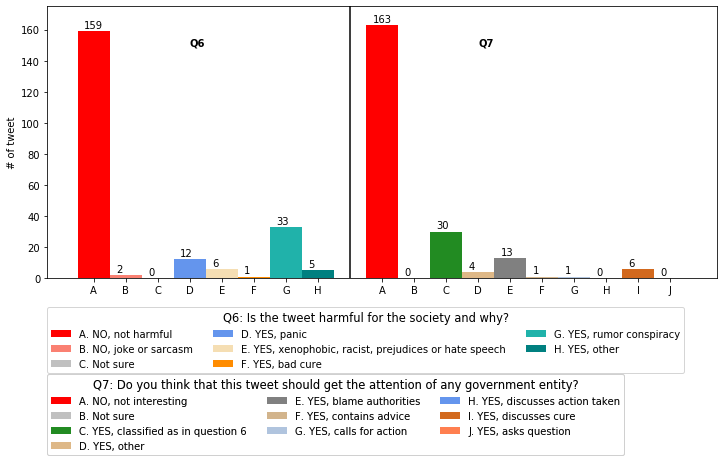}
        \caption{Questions (Q6-7).}
        \label{fig:data_dist_group2_arabic}
    \end{subfigure}
    \caption{Distribution of class labels for \textbf{Arabic tweets}}
    \label{fig:data_dist_arabic_all}
\end{figure*}

In Figure \ref{fig:data_dist_english_all} and \ref{fig:data_dist_arabic_all}, we report detailed class label distribution of each question. In general the class distributions are similar in both English and Arabic. 

\section{Experimental Parameters and Results}

\subsection{Transformers Parameters}
Below we list the hyperparameters that we used for training across all Transformers based models. All experimental scripts will be publicly available. 
\begin{itemize}
\itemsep-0.2em 
    \item Batch size: 8
    \item Learning rate (Adam): 2e-5
    \item Number of epochs: 3
    \item Max seq length: 128
\end{itemize}

\textbf{Number of parameters:}
\begin{itemize}
\itemsep-0.2em 
    \item \textbf{BERT} (bert-base-uncased): L=12, H=768, A=12, total parameters = 110M; where \textit{L} is number of layers (i.e., Transformer blocks), \textit{H} is the hidden size, and \textit{A} is the number of self-attention heads.
    \item \textbf{RoBERTa} (roberta-base): similar to BERT-base with a higher number of parameters (125M).
    \item \textbf{ALBERT} (albert-base-v1): similar to BERT-base with a reduced parameters size of 12M. 
    \item \textbf{AraBERT} (bert-base-arabert): same number as BERT (110M).
    \item \textbf{BERT Multilingual} (bert-base-multilingual-uncased) (mBERT): similar to BERT-base with a higher number of parameters (172M).
    \item \textbf{XML-RoBERTa} (xlm-roberta-base): L=12, H=768, A=12; total parameters = 270M.
\end{itemize}

\subsection{FastText Parameters}
We plan to release all the FastText parameters with our released packages. We have not listed them here due to their exhaustive list.   

\subsection{Computing Infrastructure and Runtime}
We used the NVIDIA Tesla V100-SXM2-32 GB GPU machine consists of 56 cores and 256GB CPU memory. 
To perform an experiment for a question on average the computing time took 40 minutes using a BERT base model, which results in around 4 hours for seven questions using one transformer architecture.

\subsection{Results}
\label{app-sec:results}
The detail classification results on dev and test sets in terms of accuracy (Acc), macro-F1 (M-F1) and weighted-F1 (W-F1) for English data are reported in Table \ref{tab:results_english_dev_binary_multi_labels} and \ref{tab:results_english_binary_multi_labels}, respectively.
For Arabic data the detail of dev and test sets results are reported in Table \ref{tab:results_arabic_dev_binary_multi_labels} and \ref{tab:results_arabic_binary_multi_labels}, respectively. 

In Table \ref{tab:results_english_dev_multilang} and \ref{tab:results_english_multilang}, we report on dev and test set for multilingual setting where training is performed by combining English and Arabic data and evaluated on English data. 
With the same multilingual setting the results on Arabic evaluation is reported in Table \ref{tab:results_arabic_dev_multilang} and \ref{tab:results_arabic_multilang} for dev and test sets, respectively.

\begin{table}[]
\centering
\scalebox{0.70}{
\begin{tabular}{lrrr|rrr}
\toprule
\multicolumn{1}{c}{\textbf{}} & \multicolumn{3}{c|}{\textbf{Binary}} & \multicolumn{3}{c}{\textbf{Multiclass}} \\  \midrule
\multicolumn{1}{c}{\textbf{Q.}} & \multicolumn{1}{c}{\textbf{Acc.}} & \multicolumn{1}{c}{\textbf{M-F1}} & \multicolumn{1}{c|}{\textbf{W-F1}} & \multicolumn{1}{c}{\textbf{Acc.}} & \multicolumn{1}{c}{\textbf{M-F1}} & \multicolumn{1}{c}{\textbf{W-F1}} \\  \midrule
\multicolumn{7}{c}{\textbf{Majority}} \\  \midrule
Q1 & 60.8 & 37.8 & 46.0 & - & - & - \\
Q2 & 85.2 & 46.0 & 78.4 & 51.1 & 18.9 & 44.0 \\
Q3 & 80.0 & 44.4 & 71.1 & 54.1 & 22.5 & 48.3 \\
Q4 & 58.1 & 36.7 & 42.7 & 36.4 & 27.5 & 35.5 \\
Q5 & 51.6 & 34.0 & 35.1 & 38.7 & 33.2 & 37.6 \\
Q6 & 78.4 & 44.0 & 69.0 & 63.9 & 12.60 & 53.9 \\
Q7 & 64.0 & 39.0 & 50.0 & 64.9 & 14.00 & 57.8 \\ \midrule
\multicolumn{7}{c}{\textbf{BERT}} \\ \midrule
Q1 & 92.0 & 91.5 & 91.9 & - & - & - \\
Q2 & 88.1 & 66.0 & 85.3 & 63.5 & 24.3 & 52.5 \\
Q3 & 90.3 & 81.2 & 89.1 & 58.1 & 28.1 & 54.1 \\
Q4 & 90.0 & 89.7 & 90.0 & 51.9 & 33.4 & 44.5 \\
Q5 & 89.0 & 89.0 & 89.0 & 61.9 & 51.8 & 58.0 \\
Q6 & 92.5 & 88.0 & 92.2 & 69.4 & 15.5 & 59.2 \\
Q7 & 93.4 & 92.8 & 93.4 & 62.9 & 10.70 & 57.5 \\ \midrule
\multicolumn{7}{c}{\textbf{RoBERTa}} \\  \midrule
Q1 & 93.7 & 93.4 & 93.7 & - & - & - \\
Q2 & 87.0 & 60.7 & 83.3 & 64.5 & 36.0 & 55.5 \\
Q3 & 87.3 & 76.0 & 85.9 & 54.2 & 28.7 & 52.4 \\
Q4 & 91.3 & 91.0 & 91.2 & 51.9 & 35.1 & 47.6 \\
Q5 & 85.5 & 85.4 & 85.4 & 67.7 & 56.9 & 64.1 \\
Q6 & 89.2 & 82.8 & 88.8 & 72.2 & 20.1 & 64.2 \\
Q7 & 90.8 & 89.9 & 90.7 & 61.6 & 11.40 & 57.6 \\ \midrule
\multicolumn{7}{c}{\textbf{ALBERT}} \\ \midrule
Q1 & 91.8 & 91.2 & 91.7 & - & - & - \\
Q2 & 87.8 & 66.6 & 85.3 & 60.0 & 20.2 & 47.4 \\
Q3 & 89.0 & 77.8 & 87.3 & 52.6 & 24.8 & 49.3 \\
Q4 & 85.8 & 85.4 & 85.8 & 47.1 & 31.0 & 41.8 \\
Q5 & 82.9 & 82.9 & 82.9 & 68.1 & 57.6 & 64.6 \\
Q6 & 86.3 & 74.6 & 84.4 & 71.6 & 17.6 & 62.1 \\
Q7 & 89.2 & 88.4 & 89.2 & 63.1 & 9.70 & 54.7 \\ \midrule
\multicolumn{7}{c}{\textbf{FastText}} \\ \midrule
Q1 & 74.3 & 72.2 & 73.8 & - & - & - \\
Q2 & 85.9 & 56.6 & 81.7 & 60.6 & 26.6 & 51.8 \\
Q3 & 80.0 & 55.8 & 75.4 & 55.5 & 23.2 & 48.9 \\
Q4 & 75.5 & 74.7 & 75.4 & 43.5 & 32.4 & 42.3 \\
Q5 & 67.1 & 66.6 & 66.7 & 45.2 & 41.7 & 45.2 \\
Q6 & 77.3 & 55.6 & 73.2 & 68.0 & 15.9 & 58.8 \\
Q7 & 75.6 & 71.6 & 74.6 & 68.4 & 16.5 & 61.8 \\ \bottomrule
\end{tabular}
}
\caption{Classification results on \textbf{dev set (English data)} using different models including majority baseline for different questions. Acc. - Accuracy, M-F1 - macro F1, W-F1 - weighted average F1. For Q1, binary and multiclass setting's results are same. 
}
\label{tab:results_english_dev_binary_multi_labels}
\end{table}

\begin{table}[]
\centering
\scalebox{0.70}{
\begin{tabular}{@{}lrrr|rrr@{}}
\toprule
\multicolumn{4}{c}{\textbf{Binary}} & \multicolumn{3}{|c}{\textbf{Multiclass}} \\\midrule
\multicolumn{1}{c}{\textbf{Q.}} & \multicolumn{1}{c}{\textbf{Acc.}} & \multicolumn{1}{c}{\textbf{M-F1}} & \multicolumn{1}{c}{\textbf{W-F1}} & \multicolumn{1}{|c}{\textbf{Acc.}} & \multicolumn{1}{c}{\textbf{M-F1}} & \multicolumn{1}{c}{\textbf{W-F1}} \\ \midrule
\multicolumn{7}{c}{\textbf{Majority}} \\ \midrule
Q1 & 60.5 & 37.7 & 45.6 & - & - & - \\
Q2 & 85.8 & 46.2 & 79.2 & 58.0 & 14.7 & 42.6 \\
Q3 & 81.1 & 44.8 & 72.7 & 59.0 & 14.8 & 43.8 \\
Q4 & 58.7 & 37.0 & 43.5 & 36.4 & 10.7 & 19.4 \\
Q5 & 52.5 & 34.4 & 36.1 & 38.4 & 13.9 & 21.3 \\
Q6 & 78.7 & 44.0 & 69.3 & 66.1 & 9.90 & 52.6 \\
Q7 & 64.1 & 39.0 & 50.0 & 63.3 & 7.80 & 49.1 \\\midrule
\multicolumn{7}{c}{\textbf{BERT}} \\\midrule
Q1 & 87.7 & 87.0 & 87.6 & - & - & - \\
Q2 & 89.2 & 69.0 & 86.9 & 59.7 & 21.9 & 48.5 \\
Q3 & 86.5 & 71.1 & 84.3 & 60.3 & 30.6 & 57.6 \\
Q4 & 84.2 & 83.3 & 84.0 & 49.2 & 29.5 & 41.6 \\
Q5 & 81.3 & 81.2 & 81.3 & 54.8 & 44.6 & 50.4 \\
Q6 & 87.1 & 77.9 & 86.1 & 68.3 & 14.0 & 57.2 \\
Q7 & 89.4 & 88.4 & 89.3 & 62.7 & 9.50 & 54.6 \\\midrule
\multicolumn{7}{c}{\textbf{RoBERTa}} \\\midrule
Q1 & 90.7 & 90.1 & 90.6 & - & - & - \\
Q2 & 86.9 & 57.8 & 82.9 & 58.4 & 21.3 & 46.6 \\
Q3 & 83.5 & 65.0 & 80.8 & 52.5 & 25.9 & 50.9 \\
Q4 & 83.8 & 83.2 & 83.8 & 49.2 & 31.8 & 44.1 \\
Q5 & 73.8 & 73.6 & 73.7 & 54.4 & 43.2 & 50.3 \\
Q6 & 82.3 & 70.0 & 81.0 & 67.5 & 15.6 & 58.4 \\
Q7 & 84.9 & 83.2 & 84.7 & 59.3 & 9.70 & 55.2 \\\midrule
\multicolumn{7}{c}{\textbf{ALBERT}} \\\midrule
Q1 & 86.5 & 85.8 & 86.5 & - & - & - \\
Q2 & 87.7 & 60.3 & 83.9 & 58.7 & 17.3 & 44.8 \\
Q3 & 83.8 & 61.1 & 79.6 & 50.5 & 19.6 & 45.4 \\
Q4 & 78.5 & 77.7 & 78.5 & 45.9 & 31.0 & 39.5 \\
Q5 & 72.8 & 72.6 & 72.7 & 52.1 & 41.4 & 48.0 \\
Q6 & 82.5 & 65.1 & 79.2 & 66.9 & 13.9 & 56.5 \\
Q7 & 79.3 & 76.8 & 79.0 & 61.5 & 9.80 & 53.5 \\ \midrule
\multicolumn{7}{c}{\textbf{FastText}} \\\midrule
Q1 & 73.2 & 71.1 & 72.8 & - & - & - \\
Q2 & 86.5 & 57.4 & 82.6 & 51.1 & 18.9 & 44.0 \\
Q3 & 80.5 & 58.1 & 77.2 & 54.1 & 22.5 & 48.3 \\
Q4 & 70.3 & 68.1 & 69.6 & 36.4 & 27.5 & 35.5 \\
Q5 & 63.3 & 62.9 & 63.1 & 38.7 & 33.2 & 37.6 \\
Q6 & 75.7 & 52.7 & 71.6 & 63.9 & 12.6 & 53.9 \\
Q7 & 70.5 & 66.8 & 69.9 & 64.9 & 14.0 & 57.8 \\
\bottomrule
\end{tabular}
}
\caption{Classification results on \textbf{test set (English data)} using different models including majority baseline for different questions. 
}
\label{tab:results_english_binary_multi_labels}
\end{table}

\begin{table}[]
\centering
\scalebox{0.70}{
\begin{tabular}{lrrr|rrr}
\toprule
\multicolumn{1}{c}{\textbf{}} & \multicolumn{3}{c|}{\textbf{Binary}} & \multicolumn{3}{c}{\textbf{Multiclass}} \\ \midrule
\multicolumn{1}{c}{\textbf{Q.}} & \multicolumn{1}{c}{\textbf{Acc.}} & \multicolumn{1}{c}{\textbf{M-F1}} & \multicolumn{1}{c|}{\textbf{W-F1}} & \multicolumn{1}{c}{\textbf{Acc.}} & \multicolumn{1}{c}{\textbf{M-F1}} & \multicolumn{1}{c}{\textbf{W-F1}} \\ \midrule
\multicolumn{7}{c}{\textbf{Majority}} \\ \midrule
Q1 & 63.6 & 38.9 & 49.5 & - & - & - \\
Q2 & 71.4 & 41.7 & 59.5 & 44.3 & 12.3 & 27.2 \\
Q3 & 92.9 & 48.1 & 89.4 & 54.3 & 14.1 & 38.2 \\
Q4 & 64.3 & 39.1 & 50.3 & 48.6 & 13.1 & 31.8 \\
Q5 & 53.3 & 34.8 & 37.1 & 39.3 & 14.1 & 22.2 \\
Q6 & 72.7 & 42.1 & 61.2 & 72.9 & 12.1 & 61.5 \\
Q7 & 75.0 & 42.9 & 64.3 & 74.8 & 12.2 & 64.0 \\  \midrule
\multicolumn{7}{c}{\textbf{mBERT}} \\  \midrule
Q1 & 88.2 & 87.5 & 88.3 & - & - & - \\
Q2 & 86.4 & 82.1 & 85.9 & 54.7 & 30.1 & 49.8 \\
Q3 & 84.3 & 49.9 & 85.5 & 24.0 & 15.2 & 30.0 \\
Q4 & 85.0 & 82.5 & 84.4 & 52.0 & 23.7 & 44.4 \\
Q5 & 83.3 & 83.3 & 83.4 & 74.0 & 48.5 & 65.2 \\
Q6 & 84.1 & 78.7 & 83.6 & 31.4 & 11.0 & 42.0 \\
Q7 & 85.9 & 76.1 & 83.7 & 82.3 & 18.4 & 75.5 \\ \midrule
\multicolumn{7}{c}{\textbf{AraBERT}} \\ \midrule
Q1 & 84.5 & 82.5 & 84.1 & - & - & - \\
Q2 & 84.3 & 78.6 & 83.3 & 54.0 & 24.1 & 45.7 \\
Q3 & 67.1 & 45.7 & 74.9 & 19.3 & 12.8 & 23.5 \\
Q4 & 86.4 & 84.2 & 85.9 & 56.0 & 24.2 & 51.1 \\
Q5 & 82.7 & 82.2 & 82.4 & 72.7 & 51.5 & 65.5 \\
Q6 & 86.4 & 80.0 & 85.1 & 30.0 & 10.9 & 40.7 \\
Q7 & 86.4 & 77.1 & 84.4 & 81.4 & 16.5 & 74.2 \\ \midrule
\multicolumn{7}{c}{\textbf{XML-r}} \\ \midrule
Q1 & 79.1 & 74.9 & 77.7 & - & - & - \\
Q2 & 75.7 & 57.3 & 69.3 & 47.3 & 17.0 & 34.8 \\
Q3 & 84.3 & 49.9 & 85.5 & 22.0 & 14.4 & 25.5 \\
Q4 & 76.4 & 67.6 & 72.4 & 46.0 & 15.5 & 34.1 \\
Q5 & 68.7 & 65.6 & 66.3 & 63.3 & 36.1 & 53.3 \\
Q6 & 74.1 & 48.6 & 65.0 & 37.7 & 8.8 & 43.9 \\
Q7 & 75.0 & 42.9 & 64.3 & 80.5 & 12.7 & 71.7 \\ \midrule
\multicolumn{7}{c}{\textbf{FastText}} \\ \midrule
Q1 & 73.6 & 69.1 & 72.4 & - & - & - \\
Q2 & 85.0 & 81.2 & 84.8 & 63.3 & 38.1 & 61.2 \\
Q3 & 92.9 & 48.1 & 89.4 & 84.7 & 58.8 & 84.0 \\
Q4 & 82.1 & 78.3 & 80.9 & 68.0 & 34.2 & 64.5 \\
Q5 & 80.0 & 79.9 & 80.0 & 77.3 & 66.9 & 75.2 \\
Q6 & 82.3 & 74.6 & 80.9 & 77.3 & 26.0 & 72.4 \\
Q7 & 85.5 & 76.5 & 83.7 & 81.4 & 20.4 & 76.1 \\ \bottomrule
\end{tabular}
}
\caption{Classification results on \textbf{dev set (Arabic data)} using different models including majority baseline for different questions. 
}
\label{tab:results_arabic_dev_binary_multi_labels}
\end{table}

\begin{table}[]
\centering
\scalebox{0.73}{
\begin{tabular}{@{}lrrr|rrr@{}}
\toprule
\multicolumn{1}{c}{\textbf{}} & \multicolumn{3}{c|}{\textbf{Binary}} & \multicolumn{3}{c}{\textbf{Multiclass}} \\ \midrule
\multicolumn{1}{c}{\textbf{Q.}} & \multicolumn{1}{c}{\textbf{Acc}} & \multicolumn{1}{c}{\textbf{M-F1}} & \multicolumn{1}{c}{\textbf{W-F1}} & \multicolumn{1}{c}{\textbf{Acc}} & \multicolumn{1}{c}{\textbf{M-F1}} & \multicolumn{1}{c}{\textbf{W-F1}} \\  \midrule
\multicolumn{7}{c}{\textbf{Majority}} \\  \midrule
Q1 & 64.2 & 39.1 & 50.2 & - & - & - \\ 
Q2 & 68.9 & 40.8 & 56.2 & 44.3 & 12.3 & 27.2 \\
Q3 & 95.4 & 48.8 & 93.2 & 54.3 & 14.1 & 38.2 \\
Q4 & 65.0 & 39.4 & 51.2 & 48.6 & 13.1 & 31.8 \\
Q5 & 55.0 & 35.5 & 39.0 & 39.3 & 14.1 & 22.2 \\
Q6 & 73.9 & 42.5 & 62.7 & 72.9 & 12.1 & 61.5 \\
Q7 & 74.8 & 42.8 & 64.0 & 74.8 & 12.2 & 64.0 \\  \midrule
\multicolumn{7}{c}{\textbf{mBERT}} \\ \midrule
Q1 & 88.1 & 87.0 & 88.1 & - & - & - \\
Q2 & 80.0 & 74.7 & 79.1 & 48.6 & 25.1 & 42.8 \\
Q3 & 87.0 & 51.8 & 89.2 & 22.9 & 14.5 & 27.0 \\
Q4 & 78.8 & 76.1 & 78.5 & 52.9 & 20.1 & 43.7 \\
Q5 & 76.4 & 76.2 & 76.4 & 65.7 & 44.5 & 59.0 \\
Q6 & 81.7 & 73.3 & 80.4 & 30.7 & 10.5 & 40.9 \\
Q7 & 80.7 & 69.0 & 78.5 & 75.7 & 15.4 & 66.3 \\ \midrule
\multicolumn{7}{c}{\textbf{AraBERT}} \\ \midrule
Q1 & 83.0 & 80.6 & 82.6 & - & - & - \\
Q2 & 74.1 & 63.8 & 71.1 & 50.0 & 23.2 & 42.1 \\
Q3 & 68.7 & 45.0 & 77.8 & 17.9 & 11.1 & 21.4 \\
Q4 & 81.0 & 78.0 & 80.4 & 53.6 & 20.6 & 44.9 \\
Q5 & 76.4 & 75.7 & 76.1 & 65.0 & 42.6 & 57.7 \\
Q6 & 79.8 & 68.1 & 77.3 & 28.9 & 9.8 & 38.9 \\
Q7 & 80.7 & 67.9 & 77.9 & 74.3 & 12.2 & 63.9 \\ \midrule
\multicolumn{7}{c}{\textbf{XML-r}} \\ \midrule
Q1 & 78.9 & 73.6 & 76.9 & - & - & - \\
Q2 & 69.6 & 47.2 & 60.2 & 48.6 & 19.3 & 37.4 \\
Q3 & 87.0 & 51.8 & 89.2 & 20.0 & 14.3 & 20.0 \\
Q4 & 73.0 & 63.5 & 69.0 & 45.7 & 14.7 & 34.2 \\
Q5 & 68.6 & 65.5 & 66.5 & 55.0 & 31.2 & 46.1 \\
Q6 & 74.3 & 45.9 & 64.6 & 38.5 & 9.0 & 44.5 \\
Q7 & 74.8 & 42.8 & 64.0 & 74.8 & 12.2 & 64.0 \\  \midrule
\multicolumn{7}{c}{\textbf{FastText}} \\ \midrule
Q1 & 76.6 & 73.1 & 75.8 & - & - & - \\
Q2 & 69.6 & 61.5 & 68.2 & 49.3 & 29.2 & 47.4 \\
Q3 & 95.4 & 48.8 & 93.2 & 83.6 & 64.5 & 83.1 \\
Q4 & 79.6 & 76.8 & 79.2 & 57.1 & 28.9 & 54.4 \\
Q5 & 78.6 & 78.5 & 78.6 & 78.6 & 70.6 & 77.2 \\
Q6 & 81.2 & 71.5 & 79.4 & 82.6 & 35.5 & 79.3 \\
Q7 & 77.1 & 62.5 & 74.1 & 80.3 & 33.6 & 75.7 \\ \bottomrule
\end{tabular}
}
\caption{Classification results on \textbf{test set (Arabic data)} using different models including majority baseline for different questions. 
}
\label{tab:results_arabic_binary_multi_labels}
\end{table}